\documentclass[draft]{agujournal2019}
\usepackage{url} 
\usepackage{lineno}
\usepackage[inline]{trackchanges} 
\usepackage{soul}
\usepackage{amsmath}
\draftfalse

\journalname{Geophysical Research Letters}

\begin{document}

%
%


\title{Revisit of discrete energy bands in Galilean moon's footprint tails: remote signals of particle absorption}

%
%




\authors{Fan Yang\affil{1}, Xu-Zhi Zhou\affil{1,*}, Ying Liu\affil{2,*}, Yi-Xin Sun\affil{1}, Ze-Fan Yin\affil{1}, Yi-Xin Hao\affil{3}, Zhi-Yang Liu\affil{4}, Michel Blanc\affil{4}, Jiu-Tong Zhao\affil{1}, Dong-Wen He\affil{1}, Ya-Ze Wu\affil{1}, Shan Wang\affil{1}, Chao Yue\affil{1}, Qiu-Gang Zong\affil{1,2}}


\affiliation{1}{Institute of Space Physics and Applied Technology, Peking University, Beijing, China}
\affiliation{2}{State Key Laboratory of Lunar and Planetary Sciences, Macau University of Science and Technology, Macau 999078, China}
\affiliation{3}{Max Planck Institute for Solar System Research, Katlenburg-Lindau, Germany}
\affiliation{4}{Institut de Recherche en Astrophysique et Planétologie, CNRS-Universit Paul Sabatier, Toulouse Cedex, France}



\correspondingauthor{Xu-Zhi Zhou}{xuzhi.zhou@gmail.com}
\correspondingauthor{Ying Liu}{liuying@must.edu.mo}




\begin{keypoints}
\item Juno observations of banded flux enhancements are revisited in Galilean moon's environments, raising doubts on previous interpretations. 
\item We propose an alternative interpretation that the banded structures originate from the absorption of particles during their moon encounter.
\item This model reproduces the observations of banded structures at discrete, equally-spaced velocities.
\end{keypoints}

%
%

%
%


\begin{abstract}
Recent observations from the Juno spacecraft during its transit over flux tubes of the Galilean moons have identified sharp enhancements of particle fluxes at discrete energies. These banded structures have been suspected to originate from a bounce resonance between particles and standing Alfven waves generated by the moon-magnetospheric interaction. Here, we show that predictions from the above hypothesis are inconsistent with the observations, and propose an alternative interpretation that the banded structures are remote signals of particle absorption at the moons. In this scenario, whether a particle would encounter the moon before reaching Juno depends on the number of bounce cycles it experiences within a fixed section of drift motion determined by moon-spacecraft longitudinal separation. Therefore, the absorption bands are expected to appear at discrete, equally-spaced velocities consistent with the observations. This finding improves our understanding of moon-plasma interactions and provides a potential way to evaluate the Jovian magnetospheric models.
\end{abstract}

\section*{Plain Language Summary}
The Galilean moons of Jupiter - Io, Europa, Ganymede, and Callisto, are fascinating worlds that have attracted significant research interest since their discovery in 1610. During the space era, many studies have focused on their interactions with the Jovian magnetosphere. Recent observations from the Juno spacecraft during its passage over the Galilean moons have identified enhancements of either ion or electron fluxes at discrete energy bands linearly separated in velocity. This unique feature has been attributed to the selective acceleration of particles at specific energies that can resonate with the moon-induced Alfven waves. However, predictions from this scenario are inconsistent with the observations. We thereby propose an alternative interpretation that the observed banded structures are instead discrete gaps in an overall high-flux background originating from the absorption of charged particles as they encounter the Galilean moons. In this scenario, whether a particle would encounter the moon depends on the ratio between bounce motions and drift motions, resulting in equally-spaced velocity bands. We carry out a quantitative comparison between theoretical predictions and the observational data, to show that most of the observational signatures can be reproduced by this model. The results, therefore, shed new light on our knowledge of moon-plasma interactions.

%
%

%


%
%
%
%

\section{Introduction \label{sec:intro}}
Plasma interaction with the Galilean moons (Io, Europa, Ganymede, and Callisto) plays an important role in the environment of the Jovian magnetosphere. According to current magnetohydrodynamic theories \cite{southwoodIo1980,saurOverviewMoonMagnetosphere2021}, the relative motion of the moons to the ambient plasma environment results in the excitation of Alfven waves, which could propagate towards the polar region before they are reflected at the ionosphere. These waves eventually form the Alfven wings, a standing structure that bends the field lines in the downstream direction \cite{acunaStandingAlfvenWave1981, saurThreedimensionalPlasmaSimulation1999}.

The Alfven wings and their adjacent space environments are an active region for many different processes including plasma waves \cite{sulaimanWaveParticleInteractionsAssociated2020}, particle acceleration \cite{clarkEnergeticProtonAcceleration2020,szalayProtonAccelerationIo2020}, and auroral activities \cite{muraJunoObservationsSpot2018,schlegelAlternatingEmissionFeatures2022}. For instance, bright auroral spots named the footprint tail aurora have been identified near the magnetic footprints of the Galilean moons (most significantly, Io) and their trailing tails. They are believed to originate from the acceleration of the precipitating electrons, which in turn results from the Alfvenic perturbations and their coupling with the Jovian ionosphere \cite{sulaimanWaveParticleInteractionsAssociated2020,szalayProtonAccelerationIo2020}. Indeed, spacecraft observations within the flux tubes connected to the footprint tail aurora have shown the overall flux enhancements of both electrons \cite{allegriniFirstReportElectron2020,allegriniElectronBeamsEuropa2024,szalayNewFrameworkExplain2020,rabiaEvidenceNonMonotonicBroadband2023} and ions \cite{szalayProtonAccelerationIo2020,clarkEnergeticProtonAcceleration2020}. While the broadband electron precipitation is believed to be associated with the Alfvenic perturbations, the ion acceleration has often been attributed to the existence of ion cyclotron waves \cite{clarkEnergeticProtonAcceleration2020,glocerModelingIonConic2024}. 

The Galilean moons are also known as sinks of energetic particles in the Jovian magnetosphere. The loss of particles is achieved either through their encounter with the moon's surface or atmosphere or via interactions with the moon's intrinsic electromagnetic field, if there are any \cite{saurOverviewMoonMagnetosphere2021}. Either way, the particle fluxes in the moon's wake region could be significantly reduced compared to those in the ambient plasma environments \cite{selesnickChargeStatesEnergetic2009, paranicasIoEffectEnergetic2019,paranicasEnergeticChargedParticle2024}. This observational feature, often referred to as ``microsignatures", is most significant in the immediate downstream of the moons since the wake region of depleted particles could be gradually refilled by the ambient population via various diffusion processes. Similar processes have also been studied extensively for the lunar wake either in the terrestrial magnetosphere or in the solar wind \cite{zhangThreedimensionalLunarWake2014, kimuraElectromagneticFullParticle2008, zhangAlfvenWingsLunar2016}. Moreover, recent studies on ``microsignatures" of the Saturnian moons have provided important information on the convection pattern \cite{roussosEnergeticElectronMicrosignatures2010, roussosNumericalSimulationEnergetic2013} and diffusion processes \cite{jonesEnceladusVaryingImprint2006} in Saturn's magnetosphere.

Recently, a new type of particle energy spectra has been observed in the mid-latitude region of the flux tubes connecting to the footprint tails of the Galilean moons \cite{sarkangoResonantPlasmaAcceleration2024}, which is characterized by enhanced fluxes of either ions or electrons at discrete energy bands (see Figure \ref{fig:obs}). According to \citeA{sarkangoResonantPlasmaAcceleration2024}, these bands could originate from the bounce resonance between charged particles and the standing Alfven waves in the footprint tail. In this paper, however, we show that the expected signatures based on this scenario are not consistent with the observations. 
Moreover, we propose an alternative interpretation that the observed signatures are instead discrete gaps in an overall high-flux background, and that the lower fluxes within the bands originate from the particle absorption at the Galilean moons. In the next section, we first revisit two typical events reported in \citeA{sarkangoResonantPlasmaAcceleration2024} and examine their interpretations, before we elaborate our model in Section \ref{sec:interp}.

\section{Observations Revisited \label{sec:obs}}

\begin{figure}
\noindent\includegraphics[width=0.5\textwidth]{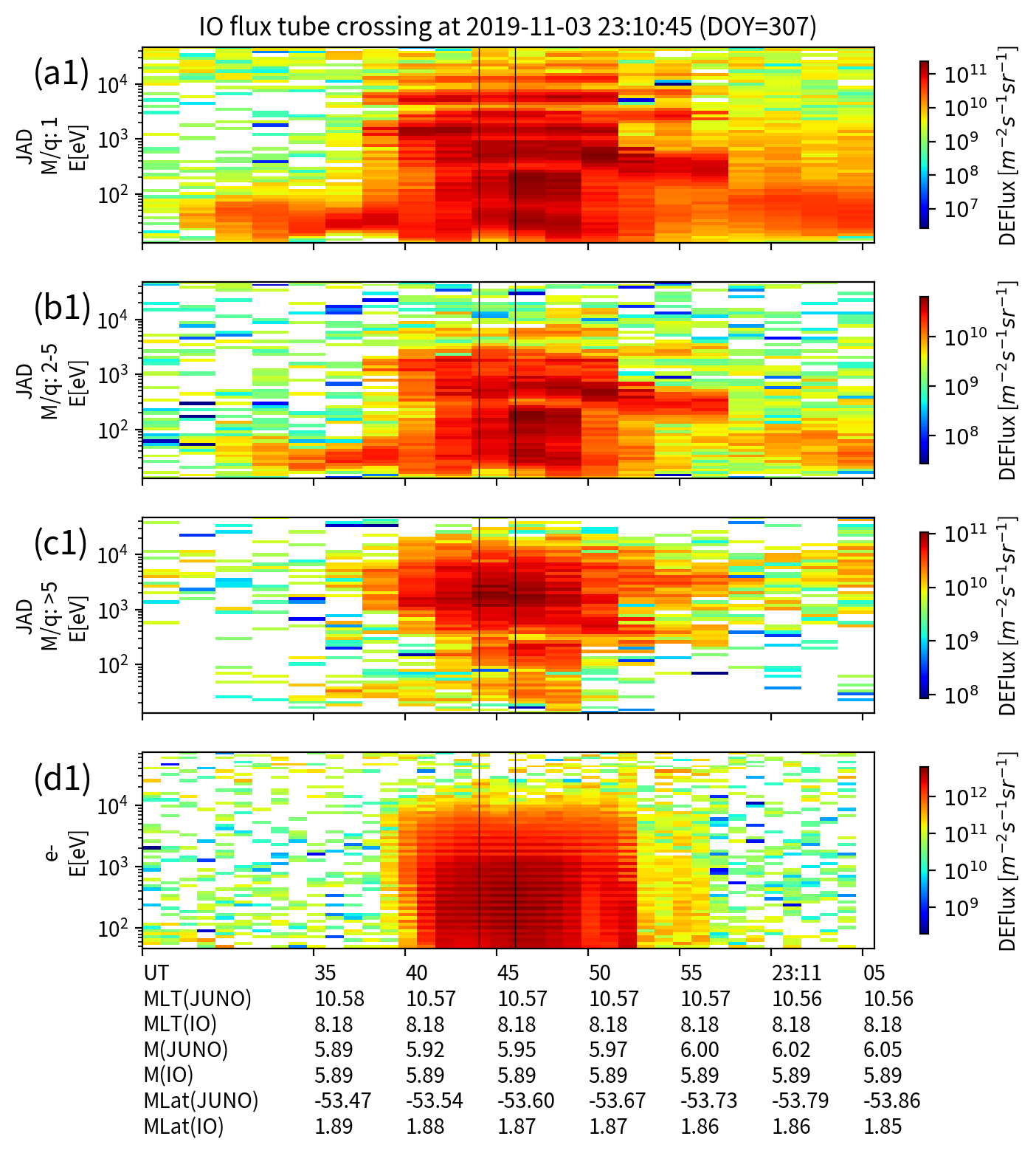}
\noindent\includegraphics[width=0.5\textwidth]{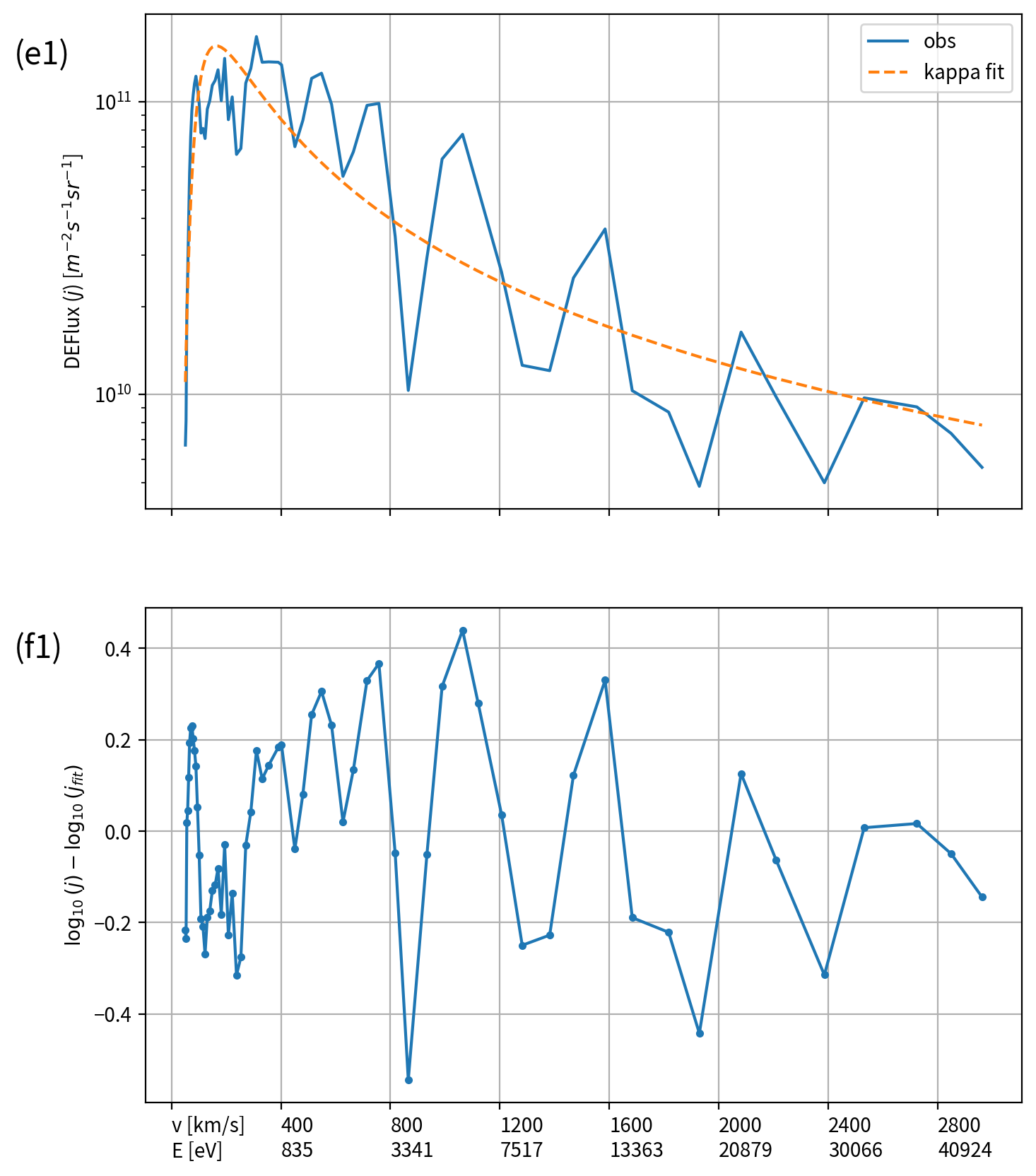}
\noindent\includegraphics[width=0.5\textwidth]{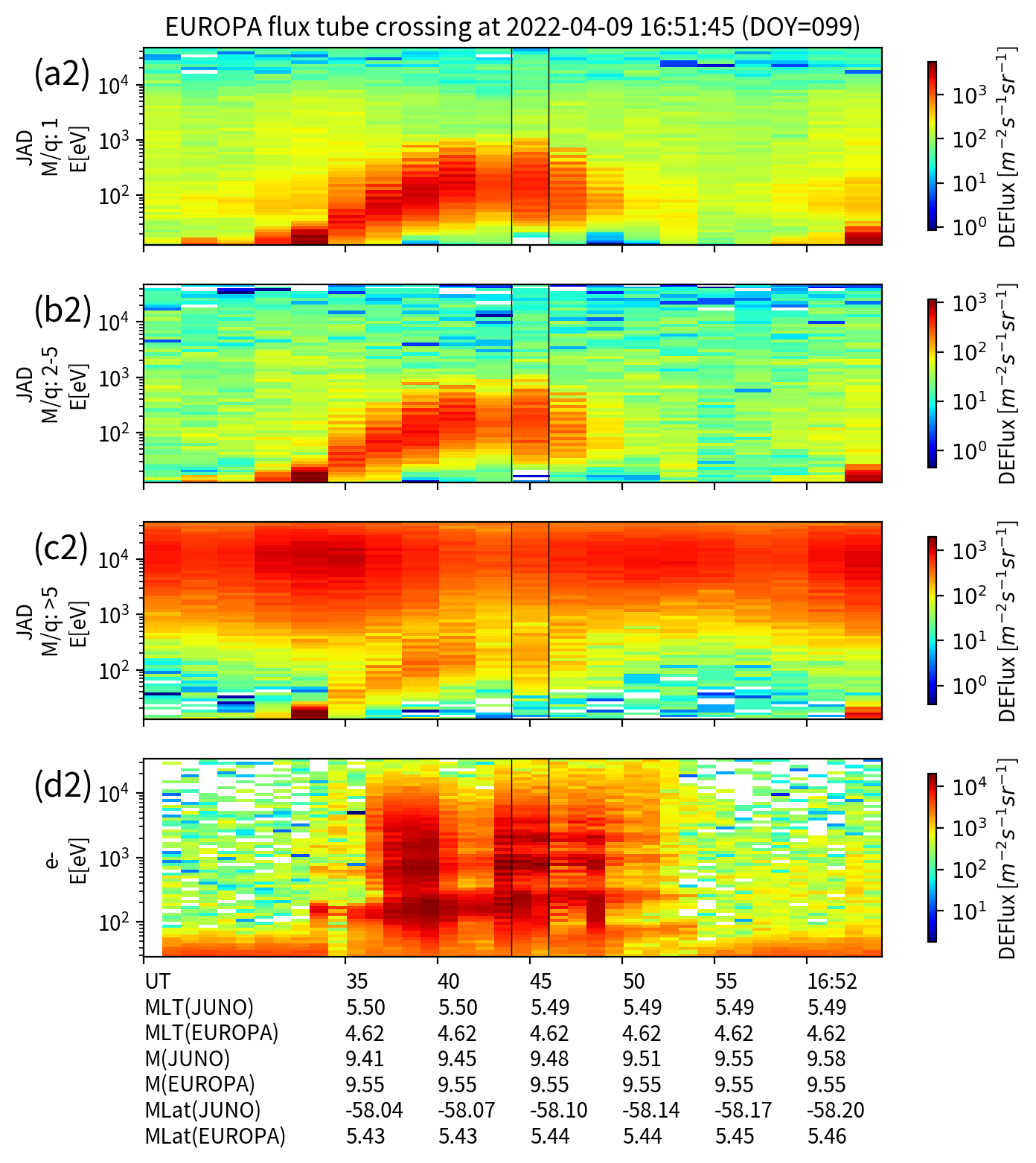}
\noindent\includegraphics[width=0.5\textwidth]{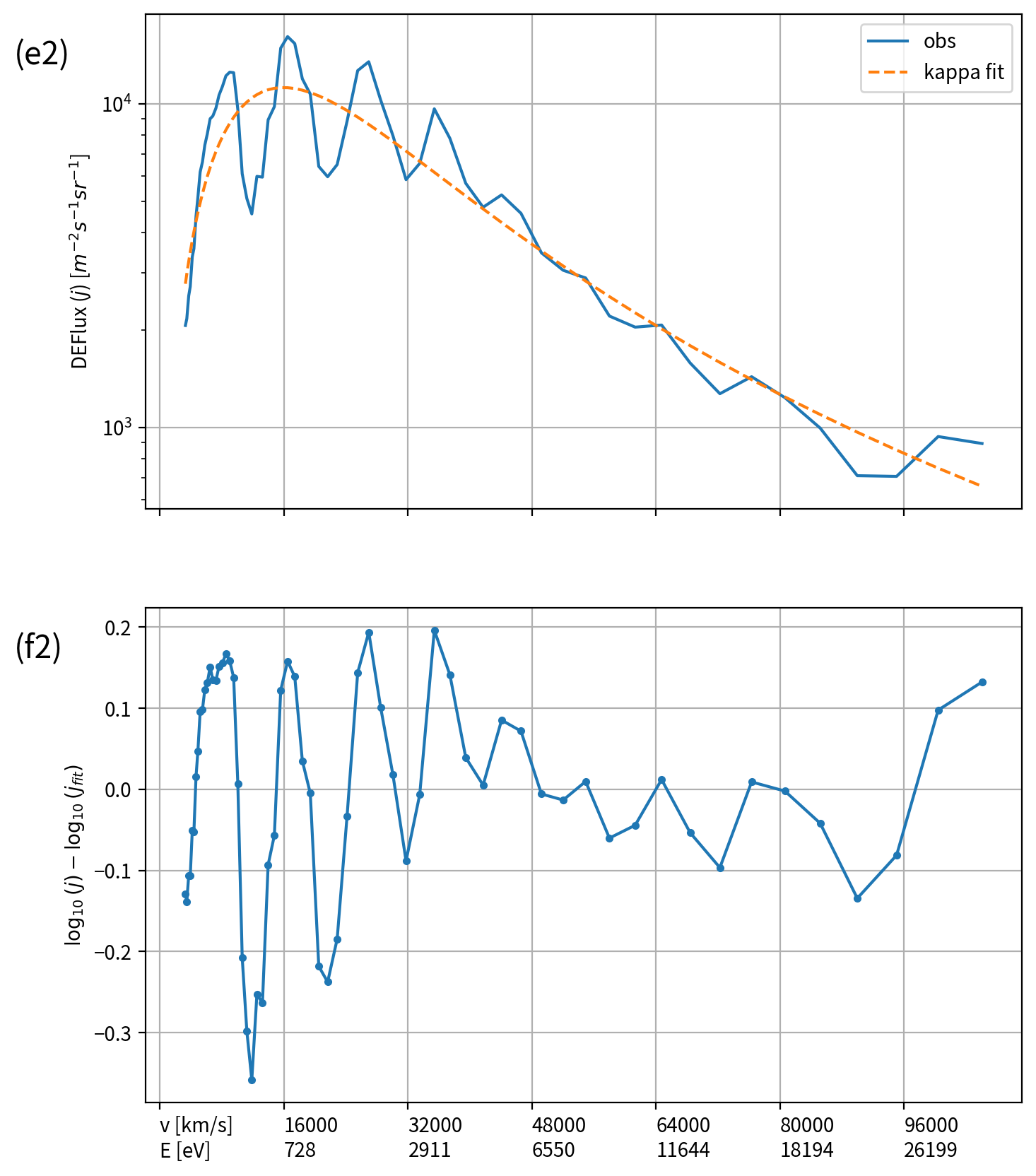}
\caption{
Juno spacecraft observations of discrete energy bands in two typical events previously reported in \citeA{sarkangoResonantPlasmaAcceleration2024}.
(upper panels) An Io flux tube crossing event at 23:10:45, 2019 DOY 307. (a1-c1) The energy spectra of ion fluxes, with $M/q$ of 1, 2-5, and $>$5, respectively. (d1) The energy spectrum of electron fluxes. (e1) The proton energy flux ($j$) as a function of proton speed, extracted from the observations in panel a1 during the central time interval bracketed by the two black lines. The red dashed line is the best fit of $j$ towards a kappa distribution, $j_{\mathrm{fit}}$. (f1) The remnant flux. (lower panels) A Europa flux tube crossing event at 16:51:45, 2022 DOY 99, in the same format as in the upper panels except that panels e2 and f2 display the electron rather than ion observations.}
\label{fig:obs}
\end{figure}

Figure \ref{fig:obs} provides the Juno spacecraft observations in the Jovian mid-latitude magnetosphere for two typical events reported in \citeA{sarkangoResonantPlasmaAcceleration2024}, the Io's flux-tube crossing event at 23:10:45, 2019 DOY 307 (the upper panels) and the Europa's flux-tube crossing event at 16:51:45, 2022 DOY 99 (lower panels). For each event, panels a-d are the observed energy spectra for three different ion species, with the mass-charge ratio $m/q$ of 1 amu/q (protons), 2-5 amu/q (likely helium or trihydrogen ions) and $>$5 amu/q (heavy ions), and for the electrons, respectively. The Juno spacecraft and the corresponding moon locations are given as tick labels below Figures \ref{fig:obs}d1 and \ref{fig:obs}d2. When deriving the M-shell and magnetic local time (MLT) locations, we adopt the JRM33 magnetic model \cite{connerneyNewModelJupiter2022} superimposed over a commonly used magnetodisc model, Con2020 \cite{connerneyJovianMagnetodiscModel2020}. Note that the MLT values, expressed through the convention of 24 hours, are calculated based on a field-line tracing to the magnetic equator. In these two events, the Juno spacecraft and the moon were both in the dawn sector, with Juno located in the mid-latitude region eastward of the moon by $\sim$2.4 and 0.88 MLT ($\sim$ 36 and 13 degrees), respectively.

In both events, the Juno spacecraft observed significant enhancements in ion and electron fluxes across a wide energy range when it reached the same M-shell as the moons. Interestingly, the strongest flux enhancements appeared at multiple, discrete energy bands for some of the particle species. In the Io crossing event (upper panels), the discrete energy bands are clearest in the proton spectrum (Figure \ref{fig:obs}a1), and they can also be vaguely seen in the energy spectra of heavier ions but not electrons. For the Europa crossing event (lower panels), clear energy bands are present in the electron spectrum (Figure \ref{fig:obs}d2) but not the ion spectra. 

To quantify the banded particle distributions, we focus on the central time interval of each event (bracketed by the vertical lines in the left panels of Figure \ref{fig:obs}) to show in Figures \ref{fig:obs}e1 and \ref{fig:obs}e2, respectively, the proton and electron energy fluxes as functions of particle speed (rather than energy, following the format of \citeA{sarkangoResonantPlasmaAcceleration2024}, see their Figure 3) in the Io and Europa crossing events. The solid lines in Figures \ref{fig:obs}e1 and \ref{fig:obs}e2 are the energy fluxes observed by Juno, $j$, and the dashed lines represent their best-fits towards a kappa distribution, $j_{\mathrm{fit}}$. We also adopt the definition of remnant fluxes in \citeA{sarkangoResonantPlasmaAcceleration2024}, $\mathrm{log}_{10}(j) - \mathrm{log}_{10}(j_{\mathrm{fit}})$, and show them in Figures \ref{fig:obs}f1 (for proton observations in the Io event) and \ref{fig:obs}f2 (for electron observations in the Europa event). In both events, the remnant fluxes display a clear, periodic trend versus particle speed.


According to \citeA{sarkangoResonantPlasmaAcceleration2024}, the occurrence of discrete energy bands in both events could originate from a bounce resonance between charged particles and low-frequency, standing Alfven waves \cite{southwoodChargedParticleBehavior1981,southwoodChargedParticleBehavior1982} so that the resonant particles could be accelerated efficiently. The resonance condition for wave-particle bounce resonance is $\omega=N\omega_b$, where $\omega_b$ represents the particle's bounce frequency, $\omega$ is the wave frequency, and $N$ is an odd or even integer for even or odd-mode standing waves, respectively. Provided a fixed wave frequency $\omega$, particles with discrete bounce frequencies, $\omega_b=\omega/N$, could resonate with the low-frequency waves. 

Also given a fixed profile of the magnetic field strength $B(s)$ along the field line in a certain M-shell, the bounce period of a particle can be expressed by

\begin{equation}
  \label{eqn:taub} 
  \tau_b = \frac{1}{v}\oint [1 - B(s)/B_m ]^{-1/2}ds = \frac{1}{v} L_s(M, \alpha_{eq}),
\end{equation}
where $B_m$ is the field strength of the particle's mirror points. The integral, $L_s$, is the spiral path length between the mirror points along the particle trajectory, which depends on the magnetic field profile of the M-shell and, less significantly, the particle's equatorial pitch angle $\alpha_{eq}$. In a simplified configuration of dipole magnetic field, the $L_s$ value could be approximated \cite{hamlinBounceFrequency1961}, by

\begin{equation}
  \label{eqn:ls} 
  L_s \simeq 4M R_J (1.3-0.56 \sin{\alpha_{eq}}),
\end{equation}
where $R_J$ is the radius of Jupiter.

Equation (\ref{eqn:taub}) indicates that the bounce period $\tau_b$ is inversely proportional to the particle speed $v$, or equivalently, the bounce frequency $\omega_b$ is proportional to $v$. Therefore, for standing Alfven waves at any given frequency $\omega$, the bounce resonance condition, $\omega=N\omega_b$, could be satisfied at discrete velocities in a harmonic sequence ($v\propto \omega / N$, i.e., $v=v_0, \frac13 v_0, \frac15v_0, \dots$ for even-mode waves and $v=\frac12 v_0, \frac14 v_0, \frac16v_0, \dots$ for odd-mode waves). This is inconsistent with the observations since the discrete energy bands occur in an arithmetic sequence instead (see Figures \ref{fig:obs}f1 and \ref{fig:obs}f2). Therefore, an alternative hypothesis is required to interpret the observational signatures, which will be proposed in the following section. 

\section{An Alternative Interpretation \label{sec:interp}}

\begin{figure}
\noindent\includegraphics[width=\textwidth]{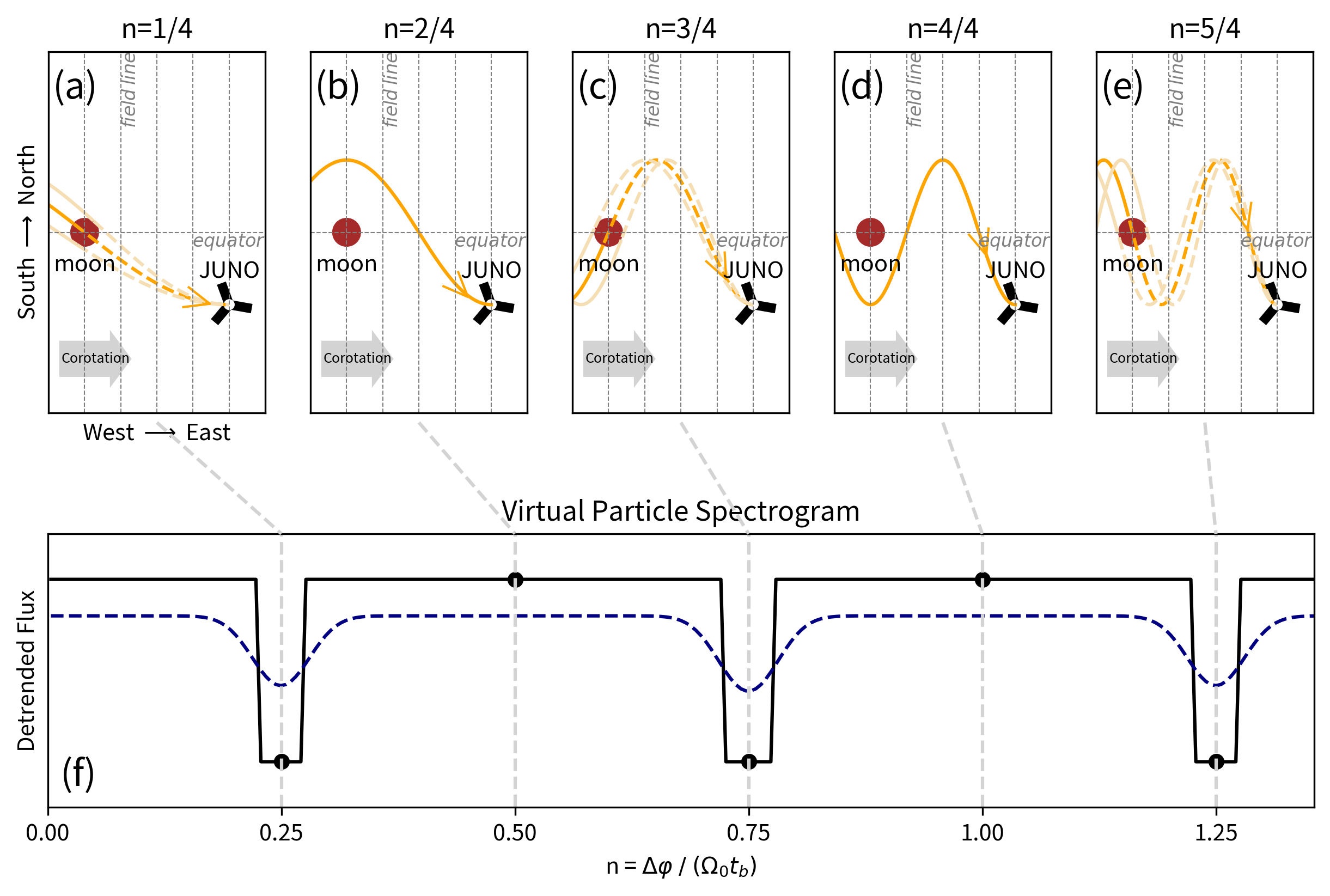}
\caption{
Schematic illustration of discrete energy bands formed by absorption of charged particles at Galilean moons. (a-e) The bounce and drift trajectories of particles at five different energies. These sample particles are chosen to experience multiple (1 to 5) quarters of bounce cycles as they drift from the moon-connecting flux tube to the Juno spacecraft. The absorption occurs in panels a, c, and e. The dashed arrows represent the hypothetical trajectories of the particles should they were not absorbed by the moon. (f) Virtual observations of particle velocity spectrum at the Juno location. The solid line shows the absorption signals at discrete, equally-spaced velocities, and the dashed line considers the diffusion effect that could occur in actual observations.}
\label{fig:cartoon}
\end{figure}

A schematic illustration of our hypothesis is given in Figure \ref{fig:cartoon}. The observational signatures are no longer interpreted as flux enhancements at discrete, resonance energies as suggested in \citeA{sarkangoResonantPlasmaAcceleration2024}, but as discrete gaps in an overall high-flux background. While the overall flux enhancements have been known to be associated with the Galilean moon's footprint tails \cite<e.g.,>{szalayProtonAccelerationIo2020,allegriniElectronBeamsEuropa2024}, the gaps are interpreted as absorption signals when the particles encounter the Galilean moons \cite{saurOverviewMoonMagnetosphere2021}. A more quantitative analysis is given below to estimate the phase-space locations of the absorption signals that could be captured in Juno observations. 

Let us consider the guiding center motion of energetic particles, which can be decomposed into drift and bounce motions (see yellow arrows in the upper panels of Figure \ref{fig:cartoon} for sample particle trajectories). Here, we only consider the region near or inside the orbit of Ganymede with M-shells lower than 15, where the energy-independent corotation drift dominates over the other drift motions (such as magnetic gradient and curvature drifts) for particles in the energy range of interest. In other words, all the particles stay in the same M-shell drifting eastward at the same angular speed. This corotation speed, $\Omega_{c}=0.633$ rad/h, is represented as the gray arrows in the upper panels of Figure \ref{fig:cartoon}. Therefore, for any given particle to drift from the moon-connecting flux tube toward Juno, the time it takes can be expressed by 

\begin{equation}
  \label{eqn:dT}
  \Delta{T} = \Delta \varphi / (\Omega_{c} - \Omega_{m}),
\end{equation}
where $\Delta \varphi$ is the longitudinal separation between the moon and the Juno spacecraft, and $\Omega_{m}$ is the Keplerian angular velocity of the moon. We should note here that $\Delta{T}$ does not depend on particle energy or species. 

For simplicity, we next focus on particles with 90$^\circ$ pitch angles when they reach the mid-latitude Juno spacecraft, and the situation of other pitch angles is left for discussion later on. In this case, the Juno spacecraft is located at the mirror point of the particle bounce motion. Since the moon is located near the magnetic equator (usually at a few-degree latitude and hence neglected below), a particle on its way to reach Juno would encounter the moon instead if $\Delta{T}$ matches an odd integer times a quarter bounce period, i.e., 

\begin{equation}
  \label{eqn:dttau}
  \Delta{T}/\tau_b = n = (2k-1)/4, \quad\quad k=1,2,3\dots, 
\end{equation}
where $n$ represents the number of bounce cycles the particle experiences during the $\Delta{T}$ interval. Figures \ref{fig:cartoon}a, \ref{fig:cartoon}c and \ref{fig:cartoon}e illustrate the cases with the $n$ number of $\frac{1}{4}$, $\frac{3}{4}$ and $\frac{5}{4}$, respectively, in which the dashed yellow arrows represent the trajectory a particle would follow should it were not absorbed by the moon. They should all correspond to the absorption signals with reduced fluxes in real spacecraft observations. In other words, the quantized $n$ numbers ($n=\frac{1}{4},\frac{3}{4},\frac{5}{4},\dots$) would indicate a series of absorption bands at discrete velocities, 

\begin{equation}
  \label{eqn:vn}
  v_n = n L_s \cdot (\Omega_{c}-\Omega_{m})/\Delta{\varphi},
\end{equation}

with the velocity separation between adjacent absorption bands given by 

\begin{equation}
  \label{eqn:dv}
  \delta{v}=L_s \cdot (\Omega_{c}-\Omega_{m})/2\Delta{\varphi}.
\end{equation}
This expectation, illustrated as the solid line in Figure \ref{fig:cartoon}f, is consistent with the observations of equally-spaced velocity bands in Figure \ref{fig:obs}.  
Based on this scenario, we also expect that the size of the moon could determine the width of the absorption bands. This effect is also illustrated in the upper panels of Figure \ref{fig:cartoon}, where particles with velocities slightly different from $v_{n}$ (see the arrows with lighter color) could also be absorbed by the moon. Given a moon radius of $R_{m}$, the upper- and lower-bound boundaries of the absorption signals in the velocity space can be expressed by

\begin{equation}
  \label{eqn:vnpm}
  v_{n\pm} = n L_s \cdot (\Omega_{c}-\Omega_{m})/(\Delta{\varphi}\mp R_m/MR_J),
\end{equation}
where $R_m/MR_J$ represents the angular size of the moon in the Jovicentric system. Their difference, or the width of the absorption bands in velocity, is then derived to be approximately

\begin{equation}
  \label{eqn:vnwidth}
  \delta v_{n\pm} \simeq 2n L_s R_m \cdot (\Omega_{c}-\Omega_{m}) /(\Delta{\varphi}^2 M R_J) = 2v_n R_m/(M R_J \Delta{\varphi}),
\end{equation}
which increases with $n$ (or equivalently, with $v_n$). Therefore, we can expect that as $n$ increases, or more specifically if $n\geq MR_J\Delta\varphi/4R_m$, the width of the absorption bands would become equal to or even larger than the band separation. In this case, the overlapped absorption bands become continuous, as no particle in this velocity range could bypass the finite-sized moon during its bounce and drift motion.  

\begin{figure}
\noindent\includegraphics[width=0.5\textwidth]{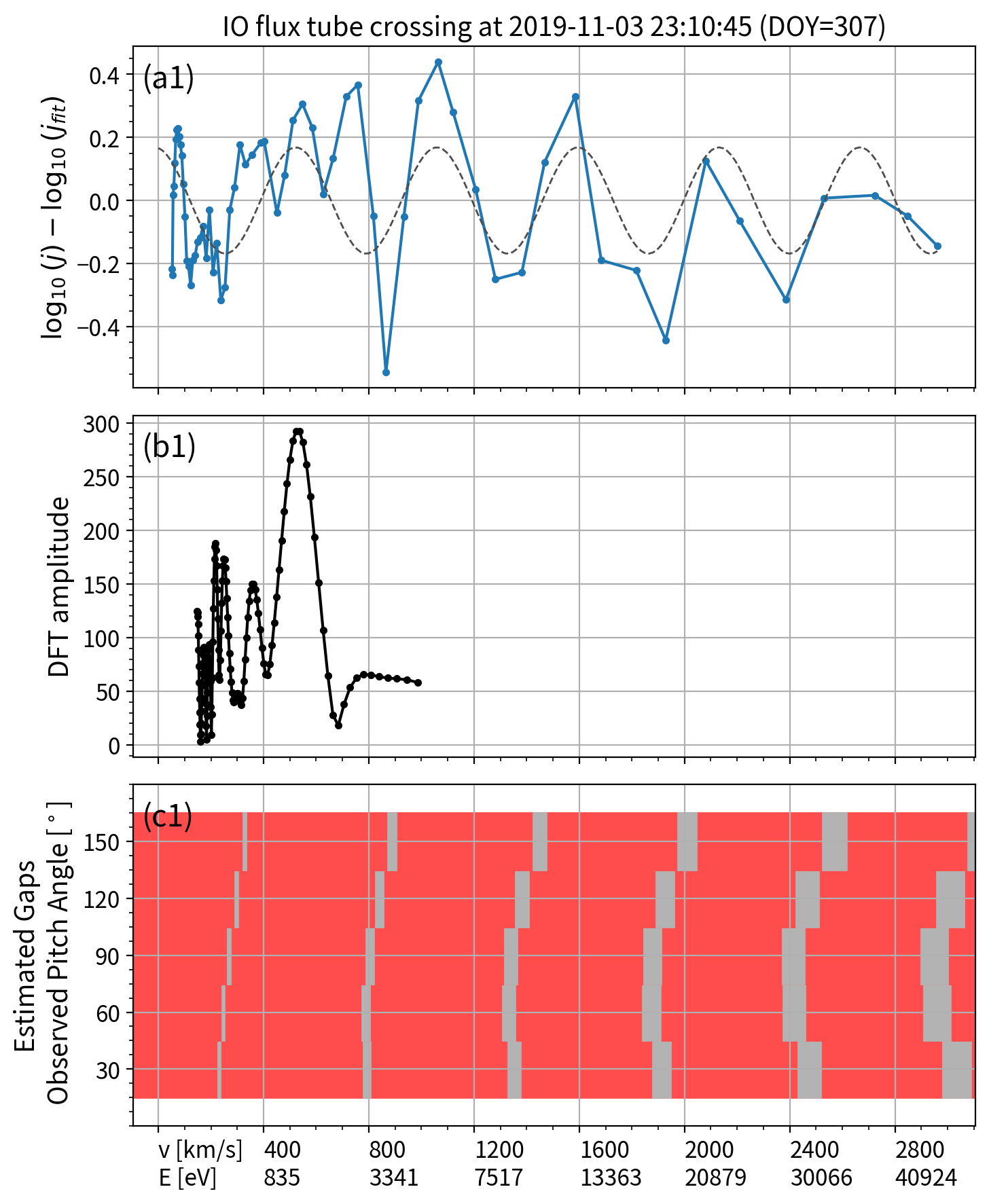}
\noindent\includegraphics[width=0.5\textwidth]{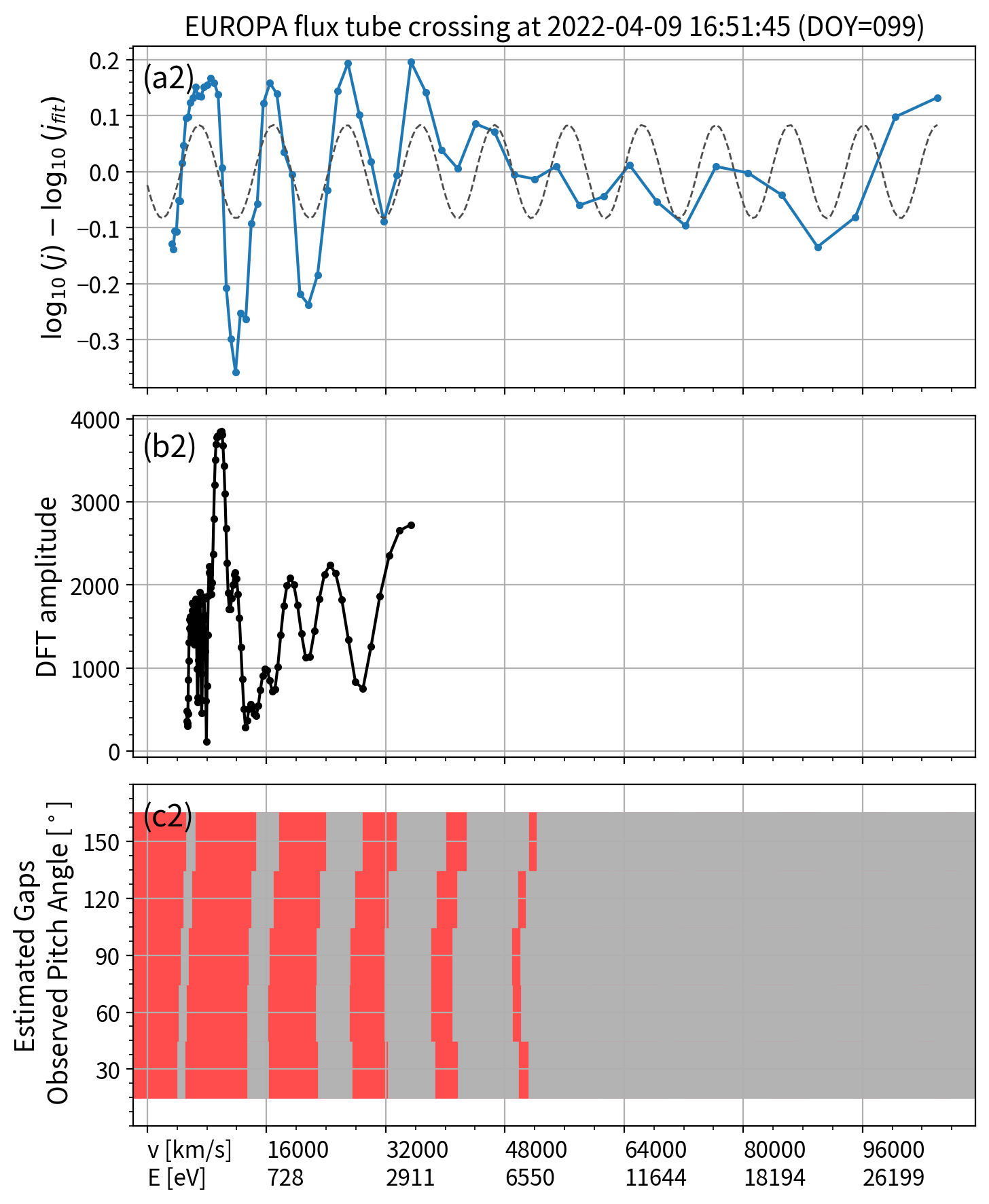}
\caption{The extraction of information, based on the hypothesis of absorption signals, from Juno observations during the Io and Europa crossing events. (a1) Remnant proton fluxes observed by Juno during its Io flux tube crossing, a replica of Figure \ref{fig:obs}e1. (b1) A discrete Fourier transform of the remnant fluxes, which shows a distinct peak at the ``frequency" of $\sim$ 540 km/s. An inverse Fourier transform of the peak ``frequency" is shown as the dashed line in panel a1. (c1) 
The estimated velocities and widths of the discrete absorption bands (the gray bands) as functions of pitch angle, embedded within the red background with higher fluxes. (a2-c2) Same analysis as in panels a1-c1, but for electron observations in the Europa flux tube crossing event. 
}
\label{fig:dft}
\end{figure}

This scenario is then applied to the observational data of the two revisited events. In Figure \ref{fig:dft}, a discrete Fourier transform (DFT) is carried out to the remnant ion or electron fluxes (the solid lines in Figures \ref{fig:dft}a1 and \ref{fig:dft}a2, which are replicas of Figures \ref{fig:obs}f1 and \ref{fig:obs}f2), to determine the major peaks in the ``frequency domain" of the particle velocity. The DFT results, shown in Figures \ref{fig:dft}b1 and \ref{fig:dft}b2, indicate the existence of a single peak in each event. One may also select the peak ``frequency" for the application of an inverse Fourier transform, to obtain a shifted cosine profile shown as the dashed lines in Figures \ref{fig:dft}a1 and \ref{fig:dft}a2, which approximately match the observations of the remnant fluxes (the solid lines). In other words, the observed banded structures are indeed equally spaced in velocity, with the separation between adjacent bands, $\delta{v}$, of $\sim$540 km/s and $\sim$10,000 km/s for the proton and electron observations in the Io and Europa crossing events, respectively. 

The $\delta{v}$ values in these two events can be substituted into Equation (\ref{eqn:dv}) to derive the longitudinal differences, $\Delta\varphi=14.5^\circ$ and $1.5^\circ$, respectively, between Juno and the corresponding moons. These values can be further substituted into Equations (\ref{eqn:vn}) and (\ref{eqn:vnwidth}) to derive the velocities and widths of the absorption bands in these two events, shown in Figures \ref{fig:dft}c1 and \ref{fig:dft}c2 as the gray areas. Here, the vertical axis provides the pitch-angle dependence of the absorption bands (to be discussed later), and for now, we focus only on the central portion at 90$^\circ$. They appear to match the observations (Figures \ref{fig:dft}a1 and \ref{fig:dft}a2) quite well, except that the observed flux variations appear to be more sinusoidal than the expectations of abrupt reduction at the boundaries of the absorption signals. This discrepancy could be explained, either by the finite energy or pitch-angle resolution of the particle data (so that particles inside and outside the absorption bands, with slightly different energies or pitch angles, could be accepted within a single energy channel) or by the occurrence of diffusion processes \cite{roussosElectronMicrodiffusionSaturnian2007} that smooths the particle spectrum (see the solid and dashed lines in Figure \ref{fig:cartoon} for an illustration). We also note that for the Europa crossing event, the discrete bands become overlapped at velocities above 50,000 km/s (see Figure \ref{fig:dft}c2), which matches the observations of much weaker flux variations above this speed (Figure \ref{fig:dft}a2). 

The vastly different $\delta{v}$ values in these two events also provide a clue for our understanding of a puzzle reported in \citeA{sarkangoResonantPlasmaAcceleration2024}, that the banded ion and electron signatures have never been observed simultaneously in a single event. According to Equation (\ref{eqn:dv}), $\delta{v}$ does not depend on particle species. However, the velocity ranges of ion and electron measurements from the Juno spacecraft differ dramatically \cite{mccomasJovianAuroralDistributions2017a}, which are approximately 45 $\sim$ 3,000 km/s and 3,250 $\sim$ 105,000 km/s, respectively. Therefore, the discrete ion bands may only be discernible if $\delta{v}$ is lower than 1,000 km/s (see the upper limit of the discrete Fourier results in Figure \ref{fig:dft}b1); however, the electron bands with such a small $\delta v$ could not be captured due to the limited energy resolution (see the lower limit of the discrete Fourier results in Figure \ref{fig:dft}b2, which is still higher than the upper limit in Figure \ref{fig:dft}b1). 

Another discrepancy between the observations and the proposed hypothesis appears in the estimation of moon-spacecraft longitudinal separation, $\Delta\varphi$. As discussed above, the $\Delta\varphi$ values inferred from the discrete Fourier transform (and a substitution to Equation (\ref{eqn:dv})) are $14.5^\circ$ and $1.5^\circ$, respectively. They both differ from the ones estimated via a field-line tracing in the modeled Jovian magnetosphere, $\sim36^\circ$ and $13^\circ$ (see Section \ref{sec:obs}). A possible explanation for this discrepancy is that the field-line tracing may not be accurate, either due to the uncertainty in the adopted Jovian magnetospheric model \cite{connerneyJovianMagnetodiscModel2020,connerneyNewModelJupiter2022} or due to the moon-induced field perturbations \cite{acunaStandingAlfvenWave1981, saurThreedimensionalPlasmaSimulation1999}. 

\begin{figure}
\noindent\includegraphics[width=0.8\textwidth]{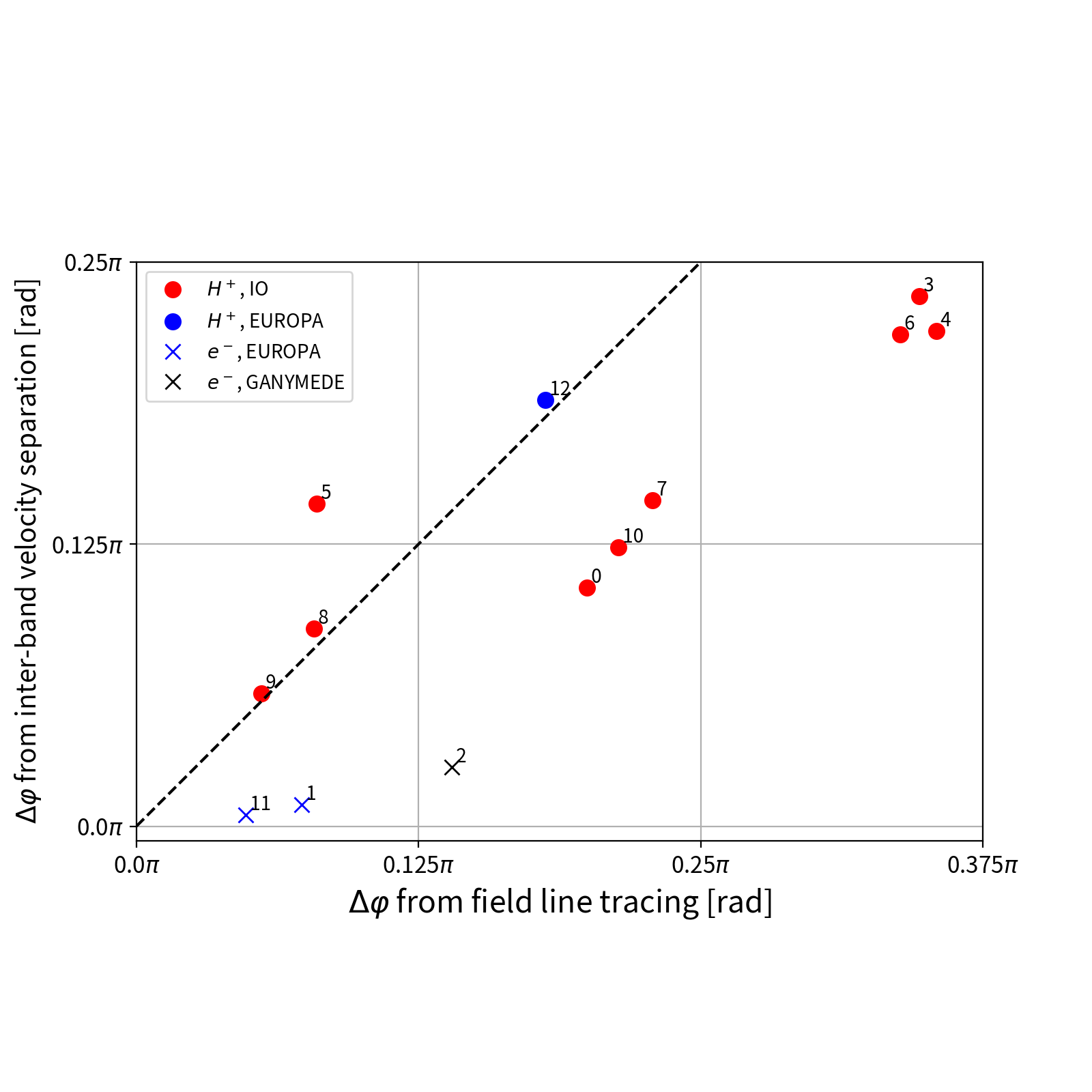}
\caption{
Statistics of $\Delta{\varphi}$ distributions based on the 13 banded events listed in Table S1. The x-axis of each data point represents the $\Delta{\varphi}$ value determined from field line tracing, and the y-axis represents the estimated $\Delta{\varphi}$ from inter-band velocity separation (based on Equation (\ref{eqn:dv})). The number next to each symbol corresponds to the event number in Table S1}
\label{fig:stats}
\end{figure}

To examine the statistical significance of the deviation, we select, from the \citeA{sarkangoResonantPlasmaAcceleration2024} event list, a total of 13 moon-crossing events with at least three discernible bands. Table S1 provides detailed information on these events. For each event in the list, we compute the longitudinal difference, $\Delta\varphi$, based on these two methods and show their comparison in Figure \ref{fig:stats}. One may find that most of the data points in the scatterplot are relatively close to the line of equality, although there is a trend that the $\Delta{\varphi}$ values given by field-line tracing are generally higher than those estimated from the Fourier transform of the discrete bands. Therefore, if we can prove that our interpretation of the discrete bands is correct, such comparisons may have the potential to help evaluate the accuracy of Jovian magnetospheric models or their applicability to the environments near the moon. This is certainly beyond the scope of our paper, and more systematic investigations may be carried out in the future.

Figure \ref{fig:stats} also displays several interesting features. The $\Delta{\varphi}$ values determined from either method are always positive, consistent with our scenario since the absorption signatures can only be expected to the east of the moon. The $\Delta{\varphi}$ values are all confined within a range below $3\pi/8$, which could be understood by the fact that the separation and width of the absorption bands are both inversely proportional to $\Delta{\varphi}$ (see Equations (\ref{eqn:dv}) and (\ref{eqn:vnwidth})). Therefore, the absorption bands at longitudinal locations far from the moons, even if not affected by diffusion, would not be discernible in the particle data with a limited energy resolution. 


\section{Conclusions and Discussion \label{sec:discuss}}

In this paper, we revisit the novel observations of discrete energy bands in particle fluxes from the Juno spacecraft in the flux tubes connected to the footprint tails of the Gailean moons. Previous studies have attributed these energy bands to bounce resonance between these particles and low-frequency waves \cite{sarkangoResonantPlasmaAcceleration2024}. However, this interpretation would lead to the enhanced fluxes in discrete velocities in a harmonic sequence, rather than in an arithmetic sequence as observed by Juno. We thereby propose an alternative interpretation that the banded structures are produced by the absorption of particles during their encounter with the Galilean moons. In this scenario, whether a particle would encounter or bypass the moon before reaching Juno depends on the number of bounce cycles it could experience within a fixed section of drift motion, which in turn is determined by the moon-spacecraft longitudinal separation. This model reproduces the banded structures at discrete, equally-spaced velocities. Moreover, most of their observational characteristics, including the velocity separation between adjacent bands and the width of each band, can be naturally explained in this scenario. 

According to our interpretation, the formation of discrete energy bands is nearly the same as the ``microsignatures" widely reported in the Saturnian and Jovian magnetospheres \cite{paranicasEnergeticChargedParticle2024}, except that the latter usually applies to particles at higher energies. In that case, the bounce frequency becomes so high that the energetic particles can only drift a short distance within a bounce cycle and therefore, cannot bypass the moon. In this sense, the ``microsignatures" are essentially the overlapped absorption bands (see the example in Figure \ref{fig:dft}c2, at velocities above 50,000 km/s). 

So far, we have only considered the signatures of particles at the pitch angle of 90$^\circ$, which indicates that the mid-latitude Juno spacecraft is located at a mirror point of their bounce motion. Particles with other pitch angles would have mirror points at higher latitudes, which indicates twofold effects on the expected observational signals. Firstly, particles with other pitch angles would have larger bounce periods $\tau_b$ (due to the dependence of $L_s$ on $\alpha_{eq}$, see Equation (\ref{eqn:ls}) for an approximation in a dipole field). Therefore, the band separation in velocity, $\delta{v}$, would also be slightly enhanced (see Equation (\ref{eqn:dv})), to cause an error in the Fourier-based estimation of $\Delta\varphi$. Secondly, the higher mirror latitudes of the particles indicate that their bounce phases would no longer be 90$^\circ$ or 270$^\circ$ when they encounter the spacecraft. Therefore, we must shift the quantized $n$ numbers for the absorption signals at pitch angles smaller or larger than 90$^\circ$. In other words, the discrete energy bands are expected to move towards lower and higher energies at pitch angles lower and higher than 90$^\circ$, respectively.

Figures \ref{fig:dft}c1 and \ref{fig:dft}c2 present the pitch-angle dependence of the absorption bands in gray colors, in which the aforementioned, twofold effects can be clearly seen. In real observations, however, these features may not be identified due to the phase mixture in omnidirectional data (like those shown in Figure \ref{fig:obs}) or in those with a low angular resolution. The phase mixing effect, as discussed before, provides another explanation for the relatively smooth variations of the observed particle fluxes near the boundaries of the energy bands. Therefore, in a future effort to evaluate the Jovian magnetospheric models (based on the comparison between observations and expectations of the discrete bands, as discussed in Section \ref{sec:interp}), it is probably important to analyze the pitch-angle resolved particle data for comparisons with the model. 

Finally, we comment on another potential source of uncertainty in our prediction of the discrete bands. In our scenario, all the particles are assumed to drift eastward at the same angular speed as Jupiter's rotation. This is a good assumption in general, since the magnetic gradient and curvature drift velocity become comparable to the corotation velocity only for particles in the MeV energy range \cite{haoFormationSaturnJupiter2020}. Also, the convection electric field in the Jovian inner magnetosphere, typically several mV/m in the dawn-dusk direction, is $\sim$ 100 times weaker than the corotation field \cite{barbosa.jovianefield1983,murakami2016}. However, it has been known that the Jovian magnetospheric rotation deviates slightly from a rigid corotation with Jupiter, especially at locations near the Io plasma torus \cite{Pontius.SubCoro1982,SzalayIonPrecipitationIO2024} due to the mass loading of the newly ionized materials \cite{Brown.massloading1994, Pontius.massloading1995}. The deviation from rigid corotation can affect our prediction on the velocity separations and widths of the absorption bands (see the $\Omega_c$ dependence in Equations (\ref{eqn:vn}) and (\ref{eqn:vnpm})), which could lead to an error in the Fourier-based estimation of $\Delta\varphi$ and thus should also be considered in our future effort of Jovian field model examination.

\section{Open Research}
The Juno observations utilized in our study, previously reported in \citeA{sarkangoResonantPlasmaAcceleration2024}, contain the data available from the NASA Planetary Data System Plasma Interactions Node, which can be found at https://doi.org/10.17189/1519715.



\acknowledgments
We thank the Juno mission and instrument teams for the excellent observational data. This study was supported by NSFC grant 42174184. 


%
%



\bibliography{ref}

\begin{thebibliography}{}

\bibitem [\protect \citeauthoryear {%
Acu{\~n}a%
, Neubauer%
\BCBL {}\ \BBA {} Ness%
}{%
Acu{\~n}a%
\ \protect \BOthers {.}}{%
{\protect \APACyear {1981}}%
}]{%
acunaStandingAlfvenWave1981}
\APACinsertmetastar {%
acunaStandingAlfvenWave1981}%
\begin{APACrefauthors}%
Acu{\~n}a, M\BPBI H.%
, Neubauer, F\BPBI M.%
\BCBL {}\ \BBA {} Ness, N\BPBI F.%
\end{APACrefauthors}%
\unskip\
\newblock
\APACrefYearMonthDay{1981}{}{}.
\newblock
{\BBOQ}\APACrefatitle {Standing {{Alfv{\'e}n}} Wave Current System at {{Io}}: {{Voyager}} 1 Observations} {Standing {{Alfv{\'e}n}} wave current system at {{Io}}: {{Voyager}} 1 observations}.{\BBCQ}
\newblock
\APACjournalVolNumPages{Journal of Geophysical Research: Space Physics}{86}{A10}{8513--8521}.
\newblock
\begin{APACrefDOI} \doi{10.1029/JA086iA10p08513} \end{APACrefDOI}
\PrintBackRefs{\CurrentBib}

\bibitem [\protect \citeauthoryear {%
Allegrini%
\ \protect \BOthers {.}}{%
Allegrini%
\ \protect \BOthers {.}}{%
{\protect \APACyear {2020}}%
}]{%
allegriniFirstReportElectron2020}
\APACinsertmetastar {%
allegriniFirstReportElectron2020}%
\begin{APACrefauthors}%
Allegrini, F.%
, Gladstone, G\BPBI R.%
, Hue, V.%
, Clark, G.%
, Szalay, J\BPBI R.%
, Kurth, W\BPBI S.%
\BDBL {}Wilson, R\BPBI J.%
\end{APACrefauthors}%
\unskip\
\newblock
\APACrefYearMonthDay{2020}{}{}.
\newblock
{\BBOQ}\APACrefatitle {First {{Report}} of {{Electron Measurements During}} a {{Europa Footprint Tail Crossing}} by {{Juno}}} {First {{Report}} of {{Electron Measurements During}} a {{Europa Footprint Tail Crossing}} by {{Juno}}}.{\BBCQ}
\newblock
\APACjournalVolNumPages{Geophysical Research Letters}{47}{18}{e2020GL089732}.
\newblock
\begin{APACrefDOI} \doi{10.1029/2020GL089732} \end{APACrefDOI}
\PrintBackRefs{\CurrentBib}

\bibitem [\protect \citeauthoryear {%
Allegrini%
\ \protect \BOthers {.}}{%
Allegrini%
\ \protect \BOthers {.}}{%
{\protect \APACyear {2024}}%
}]{%
allegriniElectronBeamsEuropa2024}
\APACinsertmetastar {%
allegriniElectronBeamsEuropa2024}%
\begin{APACrefauthors}%
Allegrini, F.%
, Saur, J.%
, Szalay, J\BPBI R.%
, Ebert, R\BPBI W.%
, Kurth, W\BPBI S.%
, Cervantes, S.%
\BDBL {}Wilson, R\BPBI J.%
\end{APACrefauthors}%
\unskip\
\newblock
\APACrefYearMonthDay{2024}{}{}.
\newblock
{\BBOQ}\APACrefatitle {Electron {{Beams}} at {{Europa}}} {Electron {{Beams}} at {{Europa}}}.{\BBCQ}
\newblock
\APACjournalVolNumPages{Geophysical Research Letters}{51}{13}{e2024GL108422}.
\newblock
\begin{APACrefDOI} \doi{10.1029/2024GL108422} \end{APACrefDOI}
\PrintBackRefs{\CurrentBib}

\bibitem [\protect \citeauthoryear {%
Barbosa%
\ \BBA {} Kivelson%
}{%
Barbosa%
\ \BBA {} Kivelson%
}{%
{\protect \APACyear {1983}}%
}]{%
barbosa.jovianefield1983}
\APACinsertmetastar {%
barbosa.jovianefield1983}%
\begin{APACrefauthors}%
Barbosa, D\BPBI D.%
\BCBT {}\ \BBA {} Kivelson, M\BPBI G.%
\end{APACrefauthors}%
\unskip\
\newblock
\APACrefYearMonthDay{1983}{}{}.
\newblock
{\BBOQ}\APACrefatitle {Dawn-dusk electric field asymmetry of the Io plasma torus} {Dawn-dusk electric field asymmetry of the io plasma torus}.{\BBCQ}
\newblock
\APACjournalVolNumPages{Geophysical Research Letters}{10}{3}{210-213}.
\newblock
\begin{APACrefDOI} \doi{https://doi.org/10.1029/GL010i003p00210} \end{APACrefDOI}
\PrintBackRefs{\CurrentBib}

\bibitem [\protect \citeauthoryear {%
Brown%
}{%
Brown%
}{%
{\protect \APACyear {1994}}%
}]{%
Brown.massloading1994}
\APACinsertmetastar {%
Brown.massloading1994}%
\begin{APACrefauthors}%
Brown, M\BPBI E.%
\end{APACrefauthors}%
\unskip\
\newblock
\APACrefYearMonthDay{1994}{}{}.
\newblock
{\BBOQ}\APACrefatitle {Observation of mass loading in the Io plasma torus} {Observation of mass loading in the io plasma torus}.{\BBCQ}
\newblock
\APACjournalVolNumPages{Geophysical Research Letters}{21}{10}{847-850}.
\newblock
\begin{APACrefDOI} \doi{https://doi.org/10.1029/94GL00564} \end{APACrefDOI}
\PrintBackRefs{\CurrentBib}

\bibitem [\protect \citeauthoryear {%
Clark%
\ \protect \BOthers {.}}{%
Clark%
\ \protect \BOthers {.}}{%
{\protect \APACyear {2020}}%
}]{%
clarkEnergeticProtonAcceleration2020}
\APACinsertmetastar {%
clarkEnergeticProtonAcceleration2020}%
\begin{APACrefauthors}%
Clark, G.%
, Mauk, B\BPBI H.%
, Kollmann, P.%
, Szalay, J\BPBI R.%
, Sulaiman, A\BPBI H.%
, Gershman, D\BPBI J.%
\BDBL {}Westlake, J.%
\end{APACrefauthors}%
\unskip\
\newblock
\APACrefYearMonthDay{2020}{}{}.
\newblock
{\BBOQ}\APACrefatitle {Energetic {{Proton Acceleration Associated With Io}}'s {{Footprint Tail}}} {Energetic {{Proton Acceleration Associated With Io}}'s {{Footprint Tail}}}.{\BBCQ}
\newblock
\APACjournalVolNumPages{Geophysical Research Letters}{47}{24}{e2020GL090839}.
\newblock
\begin{APACrefDOI} \doi{10.1029/2020GL090839} \end{APACrefDOI}
\PrintBackRefs{\CurrentBib}

\bibitem [\protect \citeauthoryear {%
Connerney%
, Timmins%
, Herceg%
\BCBL {}\ \BBA {} Joergensen%
}{%
Connerney%
\ \protect \BOthers {.}}{%
{\protect \APACyear {2020}}%
}]{%
connerneyJovianMagnetodiscModel2020}
\APACinsertmetastar {%
connerneyJovianMagnetodiscModel2020}%
\begin{APACrefauthors}%
Connerney, J\BPBI E\BPBI P.%
, Timmins, S.%
, Herceg, M.%
\BCBL {}\ \BBA {} Joergensen, J\BPBI L.%
\end{APACrefauthors}%
\unskip\
\newblock
\APACrefYearMonthDay{2020}{}{}.
\newblock
{\BBOQ}\APACrefatitle {A {{Jovian Magnetodisc Model}} for the {{Juno Era}}} {A {{Jovian Magnetodisc Model}} for the {{Juno Era}}}.{\BBCQ}
\newblock
\APACjournalVolNumPages{Journal of Geophysical Research: Space Physics}{125}{10}{e2020JA028138}.
\newblock
\begin{APACrefDOI} \doi{10.1029/2020JA028138} \end{APACrefDOI}
\PrintBackRefs{\CurrentBib}

\bibitem [\protect \citeauthoryear {%
Connerney%
\ \protect \BOthers {.}}{%
Connerney%
\ \protect \BOthers {.}}{%
{\protect \APACyear {2022}}%
}]{%
connerneyNewModelJupiter2022}
\APACinsertmetastar {%
connerneyNewModelJupiter2022}%
\begin{APACrefauthors}%
Connerney, J\BPBI E\BPBI P.%
, Timmins, S.%
, Oliversen, R\BPBI J.%
, Espley, J\BPBI R.%
, Joergensen, J\BPBI L.%
, Kotsiaros, S.%
\BDBL {}Levin, S\BPBI M.%
\end{APACrefauthors}%
\unskip\
\newblock
\APACrefYearMonthDay{2022}{}{}.
\newblock
{\BBOQ}\APACrefatitle {A {{New Model}} of {{Jupiter}}'s {{Magnetic Field}} at the {{Completion}} of {{Juno}}'s {{Prime Mission}}} {A {{New Model}} of {{Jupiter}}'s {{Magnetic Field}} at the {{Completion}} of {{Juno}}'s {{Prime Mission}}}.{\BBCQ}
\newblock
\APACjournalVolNumPages{Journal of Geophysical Research: Planets}{127}{2}{e2021JE007055}.
\newblock
\begin{APACrefDOI} \doi{10.1029/2021JE007055} \end{APACrefDOI}
\PrintBackRefs{\CurrentBib}

\bibitem [\protect \citeauthoryear {%
Glocer%
\ \protect \BOthers {.}}{%
Glocer%
\ \protect \BOthers {.}}{%
{\protect \APACyear {2024}}%
}]{%
glocerModelingIonConic2024}
\APACinsertmetastar {%
glocerModelingIonConic2024}%
\begin{APACrefauthors}%
Glocer, A.%
, {Garcia-Sage}, K.%
, Sulaiman, A.%
, Clark, G.%
, Szalay, J\BPBI R.%
, Sarkango, Y.%
\BCBL {}\ \BBA {} Bell, J.%
\end{APACrefauthors}%
\unskip\
\newblock
\APACrefYearMonthDay{2024}{}{}.
\newblock
{\BBOQ}\APACrefatitle {Modeling {{Ion Conic Formation}} in {{Io}}'s {{Auroral Footprint}}} {Modeling {{Ion Conic Formation}} in {{Io}}'s {{Auroral Footprint}}}.{\BBCQ}
\newblock
\APACjournalVolNumPages{Journal of Geophysical Research: Space Physics}{129}{4}{e2023JA032322}.
\newblock
\begin{APACrefDOI} \doi{10.1029/2023JA032322} \end{APACrefDOI}
\PrintBackRefs{\CurrentBib}

\bibitem [\protect \citeauthoryear {%
{Hamlin}%
, {Karplus}%
, {Vik}%
\BCBL {}\ \BBA {} {Watson}%
}{%
{Hamlin}%
\ \protect \BOthers {.}}{%
{\protect \APACyear {1961}}%
}]{%
hamlinBounceFrequency1961}
\APACinsertmetastar {%
hamlinBounceFrequency1961}%
\begin{APACrefauthors}%
{Hamlin}, D\BPBI A.%
, {Karplus}, R.%
, {Vik}, R\BPBI C.%
\BCBL {}\ \BBA {} {Watson}, K\BPBI M.%
\end{APACrefauthors}%
\unskip\
\newblock
\APACrefYearMonthDay{1961}{}{}.
\newblock
{\BBOQ}\APACrefatitle {{Mirror and Azimuthal Drift Frequencies for Geomagnetically Trapped Particles}} {{Mirror and Azimuthal Drift Frequencies for Geomagnetically Trapped Particles}}.{\BBCQ}
\newblock
\APACjournalVolNumPages{Journal of Geophysical Research}{66}{}{1-4}.
\newblock
\begin{APACrefDOI} \doi{10.1029/JZ066i001p00001} \end{APACrefDOI}
\PrintBackRefs{\CurrentBib}

\bibitem [\protect \citeauthoryear {%
Hao%
\ \protect \BOthers {.}}{%
Hao%
\ \protect \BOthers {.}}{%
{\protect \APACyear {2020}}%
}]{%
haoFormationSaturnJupiter2020}
\APACinsertmetastar {%
haoFormationSaturnJupiter2020}%
\begin{APACrefauthors}%
Hao, Y\BHBI X.%
, Sun, Y\BHBI X.%
, Roussos, E.%
, Liu, Y.%
, Kollmann, P.%
, Yuan, C\BHBI J.%
\BDBL {}Zong, Q\BHBI G.%
\end{APACrefauthors}%
\unskip\
\newblock
\APACrefYearMonthDay{2020}{{\APACmonth{12}}}{}.
\newblock
{\BBOQ}\APACrefatitle {The {{Formation}} of {{Saturn}}'s and {{Jupiter}}'s {{Electron Radiation Belts}} by {{Magnetospheric Electric Fields}}} {The {{Formation}} of {{Saturn}}'s and {{Jupiter}}'s {{Electron Radiation Belts}} by {{Magnetospheric Electric Fields}}}.{\BBCQ}
\newblock
\APACjournalVolNumPages{The Astrophysical Journal Letters}{905}{1}{L10}.
\newblock
\begin{APACrefDOI} \doi{10.3847/2041-8213/abca3f} \end{APACrefDOI}
\PrintBackRefs{\CurrentBib}

\bibitem [\protect \citeauthoryear {%
Jones%
\ \protect \BOthers {.}}{%
Jones%
\ \protect \BOthers {.}}{%
{\protect \APACyear {2006}}%
}]{%
jonesEnceladusVaryingImprint2006}
\APACinsertmetastar {%
jonesEnceladusVaryingImprint2006}%
\begin{APACrefauthors}%
Jones, G\BPBI H.%
, Roussos, E.%
, Krupp, N.%
, Paranicas, C.%
, Woch, J.%
, Lagg, A.%
\BDBL {}Dougherty, M\BPBI K.%
\end{APACrefauthors}%
\unskip\
\newblock
\APACrefYearMonthDay{2006}{{\APACmonth{03}}}{}.
\newblock
{\BBOQ}\APACrefatitle {Enceladus' {{Varying Imprint}} on the {{Magnetosphere}} of {{Saturn}}} {Enceladus' {{Varying Imprint}} on the {{Magnetosphere}} of {{Saturn}}}.{\BBCQ}
\newblock
\APACjournalVolNumPages{Science}{311}{5766}{1412--1415}.
\newblock
\begin{APACrefDOI} \doi{10.1126/science.1121011} \end{APACrefDOI}
\PrintBackRefs{\CurrentBib}

\bibitem [\protect \citeauthoryear {%
Kimura%
\ \BBA {} Nakagawa%
}{%
Kimura%
\ \BBA {} Nakagawa%
}{%
{\protect \APACyear {2008}}%
}]{%
kimuraElectromagneticFullParticle2008}
\APACinsertmetastar {%
kimuraElectromagneticFullParticle2008}%
\begin{APACrefauthors}%
Kimura, S.%
\BCBT {}\ \BBA {} Nakagawa, T.%
\end{APACrefauthors}%
\unskip\
\newblock
\APACrefYearMonthDay{2008}{{\APACmonth{06}}}{}.
\newblock
{\BBOQ}\APACrefatitle {Electromagnetic Full Particle Simulation of the Electric Field Structure around the Moon and the Lunar Wake} {Electromagnetic full particle simulation of the electric field structure around the moon and the lunar wake}.{\BBCQ}
\newblock
\APACjournalVolNumPages{Earth, Planets and Space}{60}{6}{591--599}.
\newblock
\begin{APACrefDOI} \doi{10.1186/BF03353122} \end{APACrefDOI}
\PrintBackRefs{\CurrentBib}

\bibitem [\protect \citeauthoryear {%
McComas%
\ \protect \BOthers {.}}{%
McComas%
\ \protect \BOthers {.}}{%
{\protect \APACyear {2017}}%
}]{%
mccomasJovianAuroralDistributions2017a}
\APACinsertmetastar {%
mccomasJovianAuroralDistributions2017a}%
\begin{APACrefauthors}%
McComas, D\BPBI J.%
, Alexander, N.%
, Allegrini, F.%
, Bagenal, F.%
, Beebe, C.%
, Clark, G.%
\BDBL {}White, D.%
\end{APACrefauthors}%
\unskip\
\newblock
\APACrefYearMonthDay{2017}{{\APACmonth{11}}}{}.
\newblock
{\BBOQ}\APACrefatitle {The {{Jovian Auroral Distributions Experiment}} ({{JADE}}) on the {{Juno Mission}} to {{Jupiter}}} {The {{Jovian Auroral Distributions Experiment}} ({{JADE}}) on the {{Juno Mission}} to {{Jupiter}}}.{\BBCQ}
\newblock
\APACjournalVolNumPages{Space Science Reviews}{213}{1-4}{547--643}.
\newblock
\begin{APACrefDOI} \doi{10.1007/s11214-013-9990-9} \end{APACrefDOI}
\PrintBackRefs{\CurrentBib}

\bibitem [\protect \citeauthoryear {%
Mura%
\ \protect \BOthers {.}}{%
Mura%
\ \protect \BOthers {.}}{%
{\protect \APACyear {2018}}%
}]{%
muraJunoObservationsSpot2018}
\APACinsertmetastar {%
muraJunoObservationsSpot2018}%
\begin{APACrefauthors}%
Mura, A.%
, Adriani, A.%
, Connerney, J\BPBI E\BPBI P.%
, Bolton, S.%
, Altieri, F.%
, Bagenal, F.%
\BDBL {}Turrini, D.%
\end{APACrefauthors}%
\unskip\
\newblock
\APACrefYearMonthDay{2018}{{\APACmonth{08}}}{}.
\newblock
{\BBOQ}\APACrefatitle {Juno Observations of Spot Structures and a Split Tail in {{Io-induced}} Aurorae on {{Jupiter}}} {Juno observations of spot structures and a split tail in {{Io-induced}} aurorae on {{Jupiter}}}.{\BBCQ}
\newblock
\APACjournalVolNumPages{Science}{361}{6404}{774--777}.
\newblock
\begin{APACrefDOI} \doi{10.1126/science.aat1450} \end{APACrefDOI}
\PrintBackRefs{\CurrentBib}

\bibitem [\protect \citeauthoryear {%
Murakami%
\ \protect \BOthers {.}}{%
Murakami%
\ \protect \BOthers {.}}{%
{\protect \APACyear {2016}}%
}]{%
murakami2016}
\APACinsertmetastar {%
murakami2016}%
\begin{APACrefauthors}%
Murakami, G.%
, Yoshioka, K.%
, Yamazaki, A.%
, Tsuchiya, F.%
, Kimura, T.%
, Tao, C.%
\BDBL {}Fujimoto, M.%
\end{APACrefauthors}%
\unskip\
\newblock
\APACrefYearMonthDay{2016}{}{}.
\newblock
{\BBOQ}\APACrefatitle {Response of Jupiter's inner magnetosphere to the solar wind derived from extreme ultraviolet monitoring of the Io plasma torus} {Response of jupiter's inner magnetosphere to the solar wind derived from extreme ultraviolet monitoring of the io plasma torus}.{\BBCQ}
\newblock
\APACjournalVolNumPages{Geophysical Research Letters}{43}{24}{12,308-12,316}.
\newblock
\begin{APACrefDOI} \doi{https://doi.org/10.1002/2016GL071675} \end{APACrefDOI}
\PrintBackRefs{\CurrentBib}

\bibitem [\protect \citeauthoryear {%
Paranicas%
\ \protect \BOthers {.}}{%
Paranicas%
\ \protect \BOthers {.}}{%
{\protect \APACyear {2019}}%
}]{%
paranicasIoEffectEnergetic2019}
\APACinsertmetastar {%
paranicasIoEffectEnergetic2019}%
\begin{APACrefauthors}%
Paranicas, C.%
, Mauk, B.%
, Haggerty, D.%
, Clark, G.%
, Kollmann, P.%
, Rymer, A.%
\BDBL {}Bolton, S\BPBI J.%
\end{APACrefauthors}%
\unskip\
\newblock
\APACrefYearMonthDay{2019}{}{}.
\newblock
{\BBOQ}\APACrefatitle {Io's {{Effect}} on {{Energetic Charged Particles}} as {{Seen}} in {{Juno Data}}} {Io's {{Effect}} on {{Energetic Charged Particles}} as {{Seen}} in {{Juno Data}}}.{\BBCQ}
\newblock
\APACjournalVolNumPages{Geophysical Research Letters}{46}{23}{13615--13620}.
\newblock
\begin{APACrefDOI} \doi{10.1029/2019GL085393} \end{APACrefDOI}
\PrintBackRefs{\CurrentBib}

\bibitem [\protect \citeauthoryear {%
Paranicas%
\ \protect \BOthers {.}}{%
Paranicas%
\ \protect \BOthers {.}}{%
{\protect \APACyear {2024}}%
}]{%
paranicasEnergeticChargedParticle2024}
\APACinsertmetastar {%
paranicasEnergeticChargedParticle2024}%
\begin{APACrefauthors}%
Paranicas, C.%
, Mauk, B\BPBI H.%
, Clark, G.%
, Kollmann, P.%
, N{\'e}non, Q.%
, Ebert, R\BPBI W.%
\BDBL {}Bolton, S.%
\end{APACrefauthors}%
\unskip\
\newblock
\APACrefYearMonthDay{2024}{}{}.
\newblock
{\BBOQ}\APACrefatitle {Energetic {{Charged Particle Measurements During Juno}}'s {{Two Close Io Flybys}}} {Energetic {{Charged Particle Measurements During Juno}}'s {{Two Close Io Flybys}}}.{\BBCQ}
\newblock
\APACjournalVolNumPages{Geophysical Research Letters}{51}{13}{e2024GL109495}.
\newblock
\begin{APACrefDOI} \doi{10.1029/2024GL109495} \end{APACrefDOI}
\PrintBackRefs{\CurrentBib}

\bibitem [\protect \citeauthoryear {%
Pontius~Jr.%
}{%
Pontius~Jr.%
}{%
{\protect \APACyear {1995}}%
}]{%
Pontius.massloading1995}
\APACinsertmetastar {%
Pontius.massloading1995}%
\begin{APACrefauthors}%
Pontius~Jr., D\BPBI H.%
\end{APACrefauthors}%
\unskip\
\newblock
\APACrefYearMonthDay{1995}{}{}.
\newblock
{\BBOQ}\APACrefatitle {Implications of variable mass loading in the Io torus: The Jovian flywheel} {Implications of variable mass loading in the io torus: The jovian flywheel}.{\BBCQ}
\newblock
\APACjournalVolNumPages{Journal of Geophysical Research: Space Physics}{100}{A10}{19531-19539}.
\newblock
\begin{APACrefDOI} \doi{https://doi.org/10.1029/95JA01554} \end{APACrefDOI}
\PrintBackRefs{\CurrentBib}

\bibitem [\protect \citeauthoryear {%
Pontius~Jr.%
\ \BBA {} Hill%
}{%
Pontius~Jr.%
\ \BBA {} Hill%
}{%
{\protect \APACyear {1982}}%
}]{%
Pontius.SubCoro1982}
\APACinsertmetastar {%
Pontius.SubCoro1982}%
\begin{APACrefauthors}%
Pontius~Jr., D\BPBI H.%
\BCBT {}\ \BBA {} Hill, T\BPBI W.%
\end{APACrefauthors}%
\unskip\
\newblock
\APACrefYearMonthDay{1982}{}{}.
\newblock
{\BBOQ}\APACrefatitle {Departure from corotation of the Io plasma torus: Local plasma production} {Departure from corotation of the io plasma torus: Local plasma production}.{\BBCQ}
\newblock
\APACjournalVolNumPages{Geophysical Research Letters}{9}{12}{1321-1324}.
\newblock
\begin{APACrefDOI} \doi{https://doi.org/10.1029/GL009i012p01321} \end{APACrefDOI}
\PrintBackRefs{\CurrentBib}

\bibitem [\protect \citeauthoryear {%
Rabia%
\ \protect \BOthers {.}}{%
Rabia%
\ \protect \BOthers {.}}{%
{\protect \APACyear {2023}}%
}]{%
rabiaEvidenceNonMonotonicBroadband2023}
\APACinsertmetastar {%
rabiaEvidenceNonMonotonicBroadband2023}%
\begin{APACrefauthors}%
Rabia, J.%
, Hue, V.%
, Szalay, J\BPBI R.%
, Andr{\'e}, N.%
, N{\'e}non, Q.%
, Blanc, M.%
\BDBL {}Sulaiman, A\BPBI H.%
\end{APACrefauthors}%
\unskip\
\newblock
\APACrefYearMonthDay{2023}{}{}.
\newblock
{\BBOQ}\APACrefatitle {Evidence for {{Non-Monotonic}} and {{Broadband Electron Distributions}} in the {{Europa Footprint Tail Revealed}} by {{Juno In Situ Measurements}}} {Evidence for {{Non-Monotonic}} and {{Broadband Electron Distributions}} in the {{Europa Footprint Tail Revealed}} by {{Juno In Situ Measurements}}}.{\BBCQ}
\newblock
\APACjournalVolNumPages{Geophysical Research Letters}{50}{12}{e2023GL103131}.
\newblock
\begin{APACrefDOI} \doi{10.1029/2023GL103131} \end{APACrefDOI}
\PrintBackRefs{\CurrentBib}

\bibitem [\protect \citeauthoryear {%
Roussos%
\ \protect \BOthers {.}}{%
Roussos%
\ \protect \BOthers {.}}{%
{\protect \APACyear {2013}}%
}]{%
roussosNumericalSimulationEnergetic2013}
\APACinsertmetastar {%
roussosNumericalSimulationEnergetic2013}%
\begin{APACrefauthors}%
Roussos, E.%
, Andriopoulou, M.%
, Krupp, N.%
, Kotova, A.%
, Paranicas, C.%
, Krimigis, S.%
\BCBL {}\ \BBA {} Mitchell, D.%
\end{APACrefauthors}%
\unskip\
\newblock
\APACrefYearMonthDay{2013}{}{}.
\newblock
{\BBOQ}\APACrefatitle {Numerical Simulation of Energetic Electron Microsignature Drifts at {{Saturn}}: {{Methods}} and Applications} {Numerical simulation of energetic electron microsignature drifts at {{Saturn}}: {{Methods}} and applications}.{\BBCQ}
\newblock
\APACjournalVolNumPages{Icarus}{226}{2}{1595--1611}.
\newblock
\begin{APACrefDOI} \doi{10.1016/j.icarus.2013.08.023} \end{APACrefDOI}
\PrintBackRefs{\CurrentBib}

\bibitem [\protect \citeauthoryear {%
Roussos%
\ \protect \BOthers {.}}{%
Roussos%
\ \protect \BOthers {.}}{%
{\protect \APACyear {2007}}%
}]{%
roussosElectronMicrodiffusionSaturnian2007}
\APACinsertmetastar {%
roussosElectronMicrodiffusionSaturnian2007}%
\begin{APACrefauthors}%
Roussos, E.%
, Jones, G\BPBI H.%
, Krupp, N.%
, Paranicas, C.%
, Mitchell, D\BPBI G.%
, Lagg, A.%
\BDBL {}Dougherty, M\BPBI K.%
\end{APACrefauthors}%
\unskip\
\newblock
\APACrefYearMonthDay{2007}{}{}.
\newblock
{\BBOQ}\APACrefatitle {Electron Microdiffusion in the {{Saturnian}} Radiation Belts: {{Cassini MIMI}}/{{LEMMS}} Observations of Energetic Electron Absorption by the Icy Moons} {Electron microdiffusion in the {{Saturnian}} radiation belts: {{Cassini MIMI}}/{{LEMMS}} observations of energetic electron absorption by the icy moons}.{\BBCQ}
\newblock
\APACjournalVolNumPages{Journal of Geophysical Research: Space Physics}{112}{A6}{}.
\newblock
\begin{APACrefDOI} \doi{10.1029/2006JA012027} \end{APACrefDOI}
\PrintBackRefs{\CurrentBib}

\bibitem [\protect \citeauthoryear {%
Roussos%
\ \protect \BOthers {.}}{%
Roussos%
\ \protect \BOthers {.}}{%
{\protect \APACyear {2010}}%
}]{%
roussosEnergeticElectronMicrosignatures2010}
\APACinsertmetastar {%
roussosEnergeticElectronMicrosignatures2010}%
\begin{APACrefauthors}%
Roussos, E.%
, Krupp, N.%
, Paranicas, C\BPBI P.%
, Mitchell, D\BPBI G.%
, M{\"u}ller, A\BPBI L.%
, Kollmann, P.%
\BDBL {}Coates, A\BPBI J.%
\end{APACrefauthors}%
\unskip\
\newblock
\APACrefYearMonthDay{2010}{}{}.
\newblock
{\BBOQ}\APACrefatitle {Energetic Electron Microsignatures as Tracers of Radial Flows and Dynamics in {{Saturn}}'s Innermost Magnetosphere} {Energetic electron microsignatures as tracers of radial flows and dynamics in {{Saturn}}'s innermost magnetosphere}.{\BBCQ}
\newblock
\APACjournalVolNumPages{Journal of Geophysical Research: Space Physics}{115}{A3}{}.
\newblock
\begin{APACrefDOI} \doi{10.1029/2009JA014808} \end{APACrefDOI}
\PrintBackRefs{\CurrentBib}

\bibitem [\protect \citeauthoryear {%
Sarkango%
\ \protect \BOthers {.}}{%
Sarkango%
\ \protect \BOthers {.}}{%
{\protect \APACyear {2024}}%
}]{%
sarkangoResonantPlasmaAcceleration2024}
\APACinsertmetastar {%
sarkangoResonantPlasmaAcceleration2024}%
\begin{APACrefauthors}%
Sarkango, Y.%
, Szalay, J\BPBI R.%
, Sulaiman, A\BPBI H.%
, Damiano, P\BPBI A.%
, McComas, D\BPBI J.%
, Rabia, J.%
\BDBL {}Allegrini, F.%
\end{APACrefauthors}%
\unskip\
\newblock
\APACrefYearMonthDay{2024}{}{}.
\newblock
{\BBOQ}\APACrefatitle {Resonant {{Plasma Acceleration}} at {{Jupiter Driven}} by {{Satellite-Magnetosphere Interactions}}} {Resonant {{Plasma Acceleration}} at {{Jupiter Driven}} by {{Satellite-Magnetosphere Interactions}}}.{\BBCQ}
\newblock
\APACjournalVolNumPages{Geophysical Research Letters}{51}{5}{e2023GL107431}.
\newblock
\begin{APACrefDOI} \doi{10.1029/2023GL107431} \end{APACrefDOI}
\PrintBackRefs{\CurrentBib}

\bibitem [\protect \citeauthoryear {%
Saur%
}{%
Saur%
}{%
{\protect \APACyear {2021}}%
}]{%
saurOverviewMoonMagnetosphere2021}
\APACinsertmetastar {%
saurOverviewMoonMagnetosphere2021}%
\begin{APACrefauthors}%
Saur, J.%
\end{APACrefauthors}%
\unskip\
\newblock
\APACrefYearMonthDay{2021}{}{}.
\newblock
{\BBOQ}\APACrefatitle {Overview of {{Moon}}--{{Magnetosphere Interactions}}} {Overview of {{Moon}}--{{Magnetosphere Interactions}}}.{\BBCQ}
\newblock
\BIn{} \APACrefbtitle {Magnetospheres in the {{Solar System}}} {Magnetospheres in the {{Solar System}}}\ (\BPGS\ 575--593).
\newblock
\APACaddressPublisher{}{American Geophysical Union (AGU)}.
\newblock
\begin{APACrefDOI} \doi{10.1002/9781119815624.ch36} \end{APACrefDOI}
\PrintBackRefs{\CurrentBib}

\bibitem [\protect \citeauthoryear {%
Saur%
, Neubauer%
, Strobel%
\BCBL {}\ \BBA {} Summers%
}{%
Saur%
\ \protect \BOthers {.}}{%
{\protect \APACyear {1999}}%
}]{%
saurThreedimensionalPlasmaSimulation1999}
\APACinsertmetastar {%
saurThreedimensionalPlasmaSimulation1999}%
\begin{APACrefauthors}%
Saur, J.%
, Neubauer, F\BPBI M.%
, Strobel, D\BPBI F.%
\BCBL {}\ \BBA {} Summers, M\BPBI E.%
\end{APACrefauthors}%
\unskip\
\newblock
\APACrefYearMonthDay{1999}{}{}.
\newblock
{\BBOQ}\APACrefatitle {Three-Dimensional Plasma Simulation of {{Io}}'s Interaction with the {{Io}} Plasma Torus: {{Asymmetric}} Plasma Flow} {Three-dimensional plasma simulation of {{Io}}'s interaction with the {{Io}} plasma torus: {{Asymmetric}} plasma flow}.{\BBCQ}
\newblock
\APACjournalVolNumPages{Journal of Geophysical Research: Space Physics}{104}{A11}{25105--25126}.
\newblock
\begin{APACrefDOI} \doi{10.1029/1999JA900304} \end{APACrefDOI}
\PrintBackRefs{\CurrentBib}

\bibitem [\protect \citeauthoryear {%
Schlegel%
\ \BBA {} Saur%
}{%
Schlegel%
\ \BBA {} Saur%
}{%
{\protect \APACyear {2022}}%
}]{%
schlegelAlternatingEmissionFeatures2022}
\APACinsertmetastar {%
schlegelAlternatingEmissionFeatures2022}%
\begin{APACrefauthors}%
Schlegel, S.%
\BCBT {}\ \BBA {} Saur, J.%
\end{APACrefauthors}%
\unskip\
\newblock
\APACrefYearMonthDay{2022}{}{}.
\newblock
{\BBOQ}\APACrefatitle {Alternating {{Emission Features}} in {{Io}}'s {{Footprint Tail}}: {{Magnetohydrodynamical Simulations}} of {{Possible Causes}}} {Alternating {{Emission Features}} in {{Io}}'s {{Footprint Tail}}: {{Magnetohydrodynamical Simulations}} of {{Possible Causes}}}.{\BBCQ}
\newblock
\APACjournalVolNumPages{Journal of Geophysical Research: Space Physics}{127}{5}{e2021JA030243}.
\newblock
\begin{APACrefDOI} \doi{10.1029/2021JA030243} \end{APACrefDOI}
\PrintBackRefs{\CurrentBib}

\bibitem [\protect \citeauthoryear {%
Selesnick%
\ \BBA {} Cohen%
}{%
Selesnick%
\ \BBA {} Cohen%
}{%
{\protect \APACyear {2009}}%
}]{%
selesnickChargeStatesEnergetic2009}
\APACinsertmetastar {%
selesnickChargeStatesEnergetic2009}%
\begin{APACrefauthors}%
Selesnick, R\BPBI S.%
\BCBT {}\ \BBA {} Cohen, C\BPBI M\BPBI S.%
\end{APACrefauthors}%
\unskip\
\newblock
\APACrefYearMonthDay{2009}{}{}.
\newblock
{\BBOQ}\APACrefatitle {Charge States of Energetic Ions in {{Jupiter}}'s Radiation Belt Inferred from Absorption Microsignatures of {{Io}}} {Charge states of energetic ions in {{Jupiter}}'s radiation belt inferred from absorption microsignatures of {{Io}}}.{\BBCQ}
\newblock
\APACjournalVolNumPages{Journal of Geophysical Research: Space Physics}{114}{A1}{}.
\newblock
\begin{APACrefDOI} \doi{10.1029/2008JA013722} \end{APACrefDOI}
\PrintBackRefs{\CurrentBib}

\bibitem [\protect \citeauthoryear {%
Southwood%
\ \BBA {} Kivelson%
}{%
Southwood%
\ \BBA {} Kivelson%
}{%
{\protect \APACyear {1981}}%
}]{%
southwoodChargedParticleBehavior1981}
\APACinsertmetastar {%
southwoodChargedParticleBehavior1981}%
\begin{APACrefauthors}%
Southwood, D\BPBI J.%
\BCBT {}\ \BBA {} Kivelson, M\BPBI G.%
\end{APACrefauthors}%
\unskip\
\newblock
\APACrefYearMonthDay{1981}{}{}.
\newblock
{\BBOQ}\APACrefatitle {Charged Particle Behavior in Low-Frequency Geomagnetic Pulsations 1. {{Transverse}} Waves} {Charged particle behavior in low-frequency geomagnetic pulsations 1. {{Transverse}} waves}.{\BBCQ}
\newblock
\APACjournalVolNumPages{Journal of Geophysical Research: Space Physics}{86}{A7}{5643--5655}.
\newblock
\begin{APACrefDOI} \doi{10.1029/JA086iA07p05643} \end{APACrefDOI}
\PrintBackRefs{\CurrentBib}

\bibitem [\protect \citeauthoryear {%
Southwood%
\ \BBA {} Kivelson%
}{%
Southwood%
\ \BBA {} Kivelson%
}{%
{\protect \APACyear {1982}}%
}]{%
southwoodChargedParticleBehavior1982}
\APACinsertmetastar {%
southwoodChargedParticleBehavior1982}%
\begin{APACrefauthors}%
Southwood, D\BPBI J.%
\BCBT {}\ \BBA {} Kivelson, M\BPBI G.%
\end{APACrefauthors}%
\unskip\
\newblock
\APACrefYearMonthDay{1982}{}{}.
\newblock
{\BBOQ}\APACrefatitle {Charged Particle Behavior in Low-Frequency Geomagnetic Pulsations, 2. {{Graphical}} Approach} {Charged particle behavior in low-frequency geomagnetic pulsations, 2. {{Graphical}} approach}.{\BBCQ}
\newblock
\APACjournalVolNumPages{Journal of Geophysical Research: Space Physics}{87}{A3}{1707--1710}.
\newblock
\begin{APACrefDOI} \doi{10.1029/JA087iA03p01707} \end{APACrefDOI}
\PrintBackRefs{\CurrentBib}

\bibitem [\protect \citeauthoryear {%
Southwood%
, Kivelson%
, Walker%
\BCBL {}\ \BBA {} Slavin%
}{%
Southwood%
\ \protect \BOthers {.}}{%
{\protect \APACyear {1980}}%
}]{%
southwoodIo1980}
\APACinsertmetastar {%
southwoodIo1980}%
\begin{APACrefauthors}%
Southwood, D\BPBI J.%
, Kivelson, M\BPBI G.%
, Walker, R\BPBI J.%
\BCBL {}\ \BBA {} Slavin, J\BPBI A.%
\end{APACrefauthors}%
\unskip\
\newblock
\APACrefYearMonthDay{1980}{}{}.
\newblock
{\BBOQ}\APACrefatitle {Io and its plasma environment} {Io and its plasma environment}.{\BBCQ}
\newblock
\APACjournalVolNumPages{Journal of Geophysical Research: Space Physics}{85}{}{5959-5968}.
\newblock
\begin{APACrefDOI} \doi{https://doi.org/10.1029/JA085iA11p05959} \end{APACrefDOI}
\PrintBackRefs{\CurrentBib}

\bibitem [\protect \citeauthoryear {%
Sulaiman%
\ \protect \BOthers {.}}{%
Sulaiman%
\ \protect \BOthers {.}}{%
{\protect \APACyear {2020}}%
}]{%
sulaimanWaveParticleInteractionsAssociated2020}
\APACinsertmetastar {%
sulaimanWaveParticleInteractionsAssociated2020}%
\begin{APACrefauthors}%
Sulaiman, A\BPBI H.%
, Hospodarsky, G\BPBI B.%
, Elliott, S\BPBI S.%
, Kurth, W\BPBI S.%
, Gurnett, D\BPBI A.%
, Imai, M.%
\BDBL {}Bolton, S\BPBI J.%
\end{APACrefauthors}%
\unskip\
\newblock
\APACrefYearMonthDay{2020}{}{}.
\newblock
{\BBOQ}\APACrefatitle {Wave-{{Particle Interactions Associated With Io}}'s {{Auroral Footprint}}: {{Evidence}} of {{Alfv{\'e}n}}, {{Ion Cyclotron}}, and {{Whistler Modes}}} {Wave-{{Particle Interactions Associated With Io}}'s {{Auroral Footprint}}: {{Evidence}} of {{Alfv{\'e}n}}, {{Ion Cyclotron}}, and {{Whistler Modes}}}.{\BBCQ}
\newblock
\APACjournalVolNumPages{Geophysical Research Letters}{47}{22}{e2020GL088432}.
\newblock
\begin{APACrefDOI} \doi{10.1029/2020GL088432} \end{APACrefDOI}
\PrintBackRefs{\CurrentBib}

\bibitem [\protect \citeauthoryear {%
Szalay%
, Allegrini%
\BCBL {}\ \protect \BOthers {.}}{%
Szalay%
, Allegrini%
\BCBL {}\ \protect \BOthers {.}}{%
{\protect \APACyear {2020}}%
}]{%
szalayNewFrameworkExplain2020}
\APACinsertmetastar {%
szalayNewFrameworkExplain2020}%
\begin{APACrefauthors}%
Szalay, J\BPBI R.%
, Allegrini, F.%
, Bagenal, F.%
, Bolton, S\BPBI J.%
, Bonfond, B.%
, Clark, G.%
\BDBL {}Wilson, R\BPBI J.%
\end{APACrefauthors}%
\unskip\
\newblock
\APACrefYearMonthDay{2020}{}{}.
\newblock
{\BBOQ}\APACrefatitle {A {{New Framework}} to {{Explain Changes}} in {{Io}}'s {{Footprint Tail Electron Fluxes}}} {A {{New Framework}} to {{Explain Changes}} in {{Io}}'s {{Footprint Tail Electron Fluxes}}}.{\BBCQ}
\newblock
\APACjournalVolNumPages{Geophysical Research Letters}{47}{18}{e2020GL089267}.
\newblock
\begin{APACrefDOI} \doi{10.1029/2020GL089267} \end{APACrefDOI}
\PrintBackRefs{\CurrentBib}

\bibitem [\protect \citeauthoryear {%
Szalay%
, Bagenal%
\BCBL {}\ \protect \BOthers {.}}{%
Szalay%
, Bagenal%
\BCBL {}\ \protect \BOthers {.}}{%
{\protect \APACyear {2020}}%
}]{%
szalayProtonAccelerationIo2020}
\APACinsertmetastar {%
szalayProtonAccelerationIo2020}%
\begin{APACrefauthors}%
Szalay, J\BPBI R.%
, Bagenal, F.%
, Allegrini, F.%
, Bonfond, B.%
, Clark, G.%
, Connerney, J\BPBI E\BPBI P.%
\BDBL {}Levin, S\BPBI M.%
\end{APACrefauthors}%
\unskip\
\newblock
\APACrefYearMonthDay{2020}{}{}.
\newblock
{\BBOQ}\APACrefatitle {Proton {{Acceleration}} by {{Io}}'s {{Alfv{\'e}nic Interaction}}} {Proton {{Acceleration}} by {{Io}}'s {{Alfv{\'e}nic Interaction}}}.{\BBCQ}
\newblock
\APACjournalVolNumPages{Journal of Geophysical Research: Space Physics}{125}{1}{e2019JA027314}.
\newblock
\begin{APACrefDOI} \doi{10.1029/2019JA027314} \end{APACrefDOI}
\PrintBackRefs{\CurrentBib}

\bibitem [\protect \citeauthoryear {%
Szalay%
\ \protect \BOthers {.}}{%
Szalay%
\ \protect \BOthers {.}}{%
{\protect \APACyear {2024}}%
}]{%
SzalayIonPrecipitationIO2024}
\APACinsertmetastar {%
SzalayIonPrecipitationIO2024}%
\begin{APACrefauthors}%
Szalay, J\BPBI R.%
, Saur, J.%
, Allegrini, F.%
, Ebert, R\BPBI W.%
, Valek, P\BPBI W.%
, Clark, G.%
\BDBL {}Wilson, R\BPBI J.%
\end{APACrefauthors}%
\unskip\
\newblock
\APACrefYearMonthDay{2024}{}{}.
\newblock
{\BBOQ}\APACrefatitle {Ion Precipitation Into Io's Poles Driven by a Strong Sub-Alfvénic Interaction} {Ion precipitation into io's poles driven by a strong sub-alfvénic interaction}.{\BBCQ}
\newblock
\APACjournalVolNumPages{Geophysical Research Letters}{51}{15}{e2024GL110205}.
\newblock
\begin{APACrefDOI} \doi{https://doi.org/10.1029/2024GL110205} \end{APACrefDOI}
\PrintBackRefs{\CurrentBib}

\bibitem [\protect \citeauthoryear {%
Zhang%
\ \protect \BOthers {.}}{%
Zhang%
\ \protect \BOthers {.}}{%
{\protect \APACyear {2014}}%
}]{%
zhangThreedimensionalLunarWake2014}
\APACinsertmetastar {%
zhangThreedimensionalLunarWake2014}%
\begin{APACrefauthors}%
Zhang, H.%
, Khurana, K\BPBI K.%
, Kivelson, M\BPBI G.%
, Angelopoulos, V.%
, Wan, W\BPBI X.%
, Liu, L\BPBI B.%
\BDBL {}Liu, W\BPBI L.%
\end{APACrefauthors}%
\unskip\
\newblock
\APACrefYearMonthDay{2014}{}{}.
\newblock
{\BBOQ}\APACrefatitle {Three-Dimensional Lunar Wake Reconstructed from {{ARTEMIS}} Data} {Three-dimensional lunar wake reconstructed from {{ARTEMIS}} data}.{\BBCQ}
\newblock
\APACjournalVolNumPages{Journal of Geophysical Research: Space Physics}{119}{7}{5220--5243}.
\newblock
\begin{APACrefDOI} \doi{10.1002/2014JA020111} \end{APACrefDOI}
\PrintBackRefs{\CurrentBib}

\bibitem [\protect \citeauthoryear {%
Zhang%
\ \protect \BOthers {.}}{%
Zhang%
\ \protect \BOthers {.}}{%
{\protect \APACyear {2016}}%
}]{%
zhangAlfvenWingsLunar2016}
\APACinsertmetastar {%
zhangAlfvenWingsLunar2016}%
\begin{APACrefauthors}%
Zhang, H.%
, Khurana, K\BPBI K.%
, Kivelson, M\BPBI G.%
, Fatemi, S.%
, Holmstr{\"o}m, M.%
, Angelopoulos, V.%
\BDBL {}Liu, W\BPBI L.%
\end{APACrefauthors}%
\unskip\
\newblock
\APACrefYearMonthDay{2016}{}{}.
\newblock
{\BBOQ}\APACrefatitle {Alfv{\'e}n Wings in the Lunar Wake: {{The}} Role of Pressure Gradients} {Alfv{\'e}n wings in the lunar wake: {{The}} role of pressure gradients}.{\BBCQ}
\newblock
\APACjournalVolNumPages{Journal of Geophysical Research: Space Physics}{121}{11}{10,698--10,711}.
\newblock
\begin{APACrefDOI} \doi{10.1002/2016JA022360} \end{APACrefDOI}
\PrintBackRefs{\CurrentBib}

\end{thebibliography}


\begin{thebibliography}{}

\end{thebibliography}

%
%
%
%
%

\end{document}


%
%


\title{Supporting Information for "Revisit of discrete energy bands in Galilean moon's footprint tails: remote signals of particle absorption"}
%
%

%
%



\authors{Fan Yang\affil{1}, Xu-Zhi Zhou\affil{1,*}, Ying Liu\affil{2,*}, Yi-Xin Sun\affil{1}, Ze-Fan Yin\affil{1}, Yi-Xin Hao\affil{3}, Zhi-Yang Liu\affil{4}, Michel Blanc\affil{4}, Jiu-Tong Zhao\affil{1}, Dong-Wen He\affil{1}, Ya-Ze Wu\affil{1}, Shan Wang\affil{1}, Chao Yue\affil{1}, Qiu-Gang Zong\affil{1,2}}


\affiliation{1}{Institute of Space Physics and Applied Technology, Peking University, Beijing, China}
\affiliation{2}{State Key Laboratory of Lunar and Planetary Sciences, Macau University of Science and Technology, Macau 999078, China}
\affiliation{3}{Max Planck Institute for Solar System Research, Katlenburg-Lindau, Germany}
\affiliation{4}{Institut de Recherche en Astrophysique et Planétologie, CNRS-Universit Paul Sabatier, Toulouse Cedex, France}


%
%

%

\begin{article}

%
%

\noindent\textbf{Contents of this file}
\begin{enumerate}
\item Tables S1
\end{enumerate}

\end{article}

\begin{table}
\label{tab:case}
\caption{List of Galilean moon-crossing events with discrete energy bands in either ion or electron measurements.}
\centering
\begin{tabular}{l l l l}
\hline
Index & UT & Specie & Moon  \\
\hline
\#0 & 2019307-23:10:45 & H$^+$ & Io \\
\#1 & 2022099-16:51:45 & e$^-$ & Europa \\
\#2 & 2021289-18:19:23 & e$^-$ & Ganymede \\
\#3 & 2018197-04:49:59 & H$^+$ & Io \\
\#4 & 2018250-01:46:08 & H$^+$ & Io \\
\#5 & 2019255-03:19:00 & H$^+$ & Io \\
\#6 & 2021052-18:23:12 & H$^+$ & Io \\
\#7 & 2021106-00:18:30 & H$^+$ & Io \\
\#8 & 2022056-01:46:12 & H$^+$ & Io \\
\#9 & 2022056-02:49:30 & H$^+$ & Io \\
\#10 & 2022272-16:37:05 & H$^+$ & Io \\
\#11 & 2021052-18:31:24 & e$^-$ & Europa \\
\#12 & 2022012-09:47:42 & H$^+$ & Europa \\
\hline
\end{tabular}
\end{table}